\documentclass[prb,twocolumn,showpacs,superscriptaddress,floatfix]{revtex4}
\usepackage{graphicx,color}

\newlength{\ldag}
\newcommand{\adag}{{a^\dagger}}
\newcommand{\ana}{{a^{\phantom\dagger}\hspace{-\ldag}}}
\newcommand{\bdag}{{b^\dagger}}
\newcommand{\anb}{{b^{\phantom\dagger}\hspace{-\ldag}}}
\newcommand{\rmi}{\mathrm{i}}
\newcommand{\rmd}{\mathrm{d}}
\newcommand{\pal}{\partial_l}
\newcommand{\pat}{\partial_t}
\newcommand{\tS}{\tilde{S}}
\newcommand{\bare}{^{(\mathrm{B})}}
\newcommand{\renorm}{^{(\mathrm{R})}}
\newcommand{\ve}{\varepsilon}
\newcommand{\sgn}{\mathrm{sgn}}

\def\nbN{\ensuremath{\mathrm{I\!N}}} 

\begin{document}

\title{Continuous unitary transformations and finite-size scaling exponents in the Lipkin-Meshkov-Glick model}

\author{S\'ebastien Dusuel}
\email{sdusuel@thp.uni-koeln.de}
\affiliation{Institut f\"ur Theoretische Physik, Universit\"at zu
K\"oln, Z\"ulpicher Str. 77, 50937 K\"oln, Germany}

\author{Julien Vidal}
\email{vidal@lptmc.jussieu.fr}
\affiliation{Laboratoire de Physique Th\'eorique de la Mati\`ere Condens\'ee, CNRS UMR 7600,
Universit\'e Pierre et Marie Curie, 4 Place Jussieu, 75252 Paris Cedex 05, France}

\begin{abstract}

We analyze the finite-size scaling exponents in the  Lipkin-Meshkov-Glick model
by means of the Holstein-Primakoff representation of the spin operators and the
continuous unitary transformations method. 
This combination allows us to  exactly compute the leading corrections 
to the ground-state energy, the gap, the magnetization, and the 
two-spin correlation functions. We also present numerical calculations 
for large system
size which confirm the validity of this approach. 
Finally, we use these results to discuss the entanglement properties 
of the ground state focusing on the (rescaled)
concurrence that we compute in the thermodynamical limit.

\end{abstract}

\maketitle

%
%
\section{Introduction}
%
%
\label{sec:intro}

A growing interest has recently been devoted to the study of quantum phase
transitions in spin systems especially from the entanglement point of view. The
ground-state intricated structure has indeed been shown to be strongly affected by the
existence of a critical point as initially illustrated in the one-dimensional $(1D)$ Ising
model under magnetic field \cite{Osterloh02,Osborne02}. Following these pioneering works,
many systems have been studied and have revealed a specific behavior of the entanglement 
as measured either by the concurrence
\cite{Bose02,Alcaraz03,Osenda03,Syljuasen03_1,Syljuasen03_2,Glaser03,Gu03_1,
Lambert04,Lambert05,Stauber04,Verstraete04_1,Verstraete04_2,Stelmachovic04,Wellard04,Reslen04}  or by the entropy
\cite{Latorre03,Latorre04_1,Jin04,Latorre04_3,Korepin03,Fan04,Its04}.
In this context, the Lipkin-Meshkov-Glick (LMG) model 
\cite{Lipkin65,Meshkov65,Glick65} has focused
much attention because of its apparent simplicity. Introduced fourty years ago in nuclear physics,
this model provides a simple description of the tunneling of bosons between two degenerate levels
and can thus be used to describe many physical systems such as two-mode Bose-Einstein condensates
\cite{Cirac98} or Josephson junctions. In the thermodynamical limit, its phase diagram can be 
simply established using a semiclassical (mean-field) approach \cite{Botet82,Botet83}. 
However, at finite size, the situation is more complicated and crucial
to understand the entanglement properties of this model
\cite{Vidal04_1,Vidal04_2,Vidal04_3,Latorre04_1,Dusuel04_3}. 
A possible way to investigate this problem is to use the exact solution provided by 
the algebraic Bethe ansatz \cite{Pan99,Links03}, but it would require the computation 
of the correlation functions which is  a rather difficult task with this formalism. 

In the present work, we present an alternative route which relies on a 
combination of two well-known methods, and which we have already 
briefly sketched in Ref.~\onlinecite{Dusuel04_3}. 
First, we use the Holstein-Primakoff representation which allows us 
to write a 
$1/N$ expansion of the spin operators at arbitrary order, $N$ being the number of degrees of freedom. 
Usually, such a development is often restricted to the first order for which 
the Hamiltonian remains quadratic and can thus be easily diagonalized. 
We show that this first order calculation does not provide us with enough 
information. This implies one has to take more terms in the $1/N$ expansion, 
and thus to deal with a nonquadratic Hamiltonian. 
Here, we use the continuous unitary transformations (CUTs)
technique proposed by Wegner\cite{Wegner94} and independently by G{\l}azek and
Wilson\cite{Glazek93,Glazek94}, to compute the next orders. 
The structure of this expansion contains all relevant informations to extract the finite size
corrections of various observables at the critical point. We consider in
particular the ground-state energy, the gap, the magnetization, and the spin-spin correlation
function for which we obtain nontrivial scaling exponents. 
These results corroborate several numerical
studies \cite{Botet83,Reslen04} in which some of these exponents have already  been computed. 

In addition to giving details of the calculations yielding the results announced in
Ref.~\onlinecite{Dusuel04_3}, we present a numerical study for all the finite-size scaling exponents which corroborates the analytical predictions. Furthermore, we  investigate the broken phase whereas in Ref.~\onlinecite{Dusuel04_3}, we only considered the symmetric one. Finally, we  compute  the leading order two-spin entanglement in the 
thermodynamical limit for any anisotropy parameter.

This paper is organized as follows. In Sec.~\ref{sec:LMGmodel}, we introduce the LMG model, discuss its
symmetries and give its simple solution for the isotropic case. 
In Sec.~\ref{sec:semi_classical}, we recall 
the variational approach which constitutes the zeroth order of the $1/N$ expansion but which is sufficient
to establish the phase diagram in the thermodynamical limit. Section~\ref{sec:firstQcorrections} is devoted to
the first order corrections that are easily computed via the standard Bogoliubov transformation. In the
next section (Sec. \ref{sec:higherQcorrections}) we introduce the basics of the CUT formalism and 
apply it to the LMG Hamiltonian. These corrections allow us to give analytical expressions of the
finite-size scaling exponents. In Sec.~\ref{sec:numerics}, we present a  numerical study  that
confirms our results and sheds light on the discrepancy with previous numerics. Then, in Sec.~\ref{sec:entanglement}
we discuss the entanglement properties of the ground state focusing on the so-called concurrence which measures
the two-spin entanglement. We show, in particular, that it displays a cusp at the transition point contrary to
what is observed in $1D$ systems. Technical details are given in Appendix. 

%
%
\section{The Lipkin-Meshkov-Glick model}
%
%
\label{sec:LMGmodel}
%
\subsection{The Hamiltonian and its symmetries}
%
\label{sec:sub:ham_and_sym}

The LMG model \cite{Lipkin65,Meshkov65,Glick65} describes a set of $N$ spins half mutually interacting in the
(anisotropic) $x-y$ plane embedded in a perpendicular magnetic field pointing in the $z$ direction. The
corresponding Hamiltonian is written 
%
%
\begin{eqnarray}
\label{eq:ham1}
H&=&-\frac{\lambda}{N}\sum_{i<j}
\left( \sigma_{x}^{i}\sigma_{x}^{j} +\gamma\sigma_{y}^{i}\sigma_{y}^{j} \right)
      -h \sum_{i}\sigma_{z}^{i}, \\
\label{ham2}
&=&-\frac{2 \lambda}{N} \left(S_x^2 +\gamma S_y^2 \right) -2 h S_z +
{\lambda \over 2} (1+\gamma), \\
&=& - {\lambda \over N} (1+\gamma) \left({\bf S}^2-S_z^2 -N/2 \right)
-2h S_z \nonumber \\
\label{eq:ham3}
&&  - {\lambda \over 2 N} (1-\gamma)\left(S_+^2+S_-^2\right),
\end{eqnarray}
%
%
where the $\sigma_{\alpha}$'s are the Pauli matrices,
$S_{\alpha}=\sum_{i} \sigma_{\alpha}^{i}/2$, and
$S_\pm=S_x\pm \mathrm{i} S_y$. The prefactor $1/N$ is necessary to obtain a
finite free energy per spin in the
thermodynamical limit. For any anisotropy parameter $\gamma$, $H$
preserves the magnitude of the total spin and does not couple states
having a different number of spins pointing in the field
direction (spin-flip symmetry), namely
%
%
\begin{equation}
\left[H,{\bf S}^2 \right]=0 \quad \mbox{and} \quad
\left[H,\prod_i \sigma_z^i \right]=0.
\label{eq:symmetries}
\end{equation}
%
%

An important consequence of the spin-flip symmetry is that any
eigenstates of $\prod_i \sigma_z^i$
satisfy
%
%
\begin{equation}
\langle S_x \rangle=\langle S_y \rangle=0
\end{equation}
%
%
and
%
%
\begin{eqnarray}
\langle S_x S_z\rangle&=&\langle S_z S_x \rangle=0 ,\\
\langle S_y S_z\rangle&=&\langle S_z S_y \rangle=0.
\end{eqnarray}
%
%

In the following, we only consider a ferromagnetic dominant  coupling
$(\lambda>0, |\gamma| \leq 1)$ and without loss of generality, we set
$\lambda=1$. For a discussion of the antiferromagnetic case, we refer
the reader to Ref.~\onlinecite{Vidal04_2}.
We also restrict our discussion to the region $h \geq 0$ since the spectrum is
even under the transformation
$h\leftrightarrow -h$. In addition, we only consider the maximum spin
sector $S=N/2$ which contains the
low-energy states.

The ground-state properties can be easily studied in the thermodynamical
limit using a mean-field analysis that we briefly recall in the following
section. It is worth however to first have a look at the isotropic model for
which a complete analytical solution is available.

\begin{figure}[t]
  \centering
  \includegraphics[width=5.5cm]{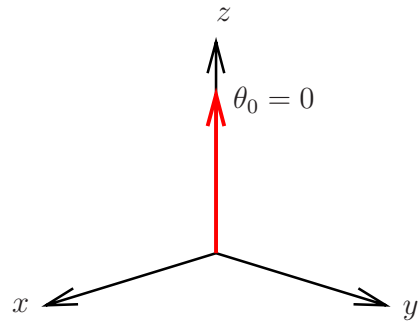}
  \caption{
    The mean-field ground state in the symmetric phase,
    for all values of the anisotropy parameter $\gamma$.}
  \label{fig:GS_sym}
\end{figure}
%

%
\subsection{Analytical solution of the isotropic model}
%
\label{sec:sub:isotropic}

In the isotropic case $\gamma=1$, the Hamiltonian (\ref{eq:ham3}) reads
%
\begin{equation}
      \label{eq:ham_isotropic}
      H=- {2\over N} \left({\bf S}^2-S_z^2 - {N \over 2} \right) -2h S_z.
\end{equation}
%
$H$ thus commutes both with ${\bf S}^2$ and $S_z$, so that it is diagonal in
the standard eigenbasis $\left\{|S,M\rangle \right\}$ of ${\bf S}^2$ and
$S_z$. The eigenenergies are
%
\begin{equation}
      \label{eq:energies_isotropic}
      E(S,M) = - {2\over N} \left[S(S+1)-M^2 -N/2 \right] -2h M.
\end{equation}
%
For the ferromagnetic case that we consider here, the ground state
is then obtained for $S=N/2$ and
%
%
\begin{equation}
M_0=\Bigg\{
\begin{array}{ccc}
I(h N /2)& {\rm for}&  0\leq h < 1 \\
N/2 & {\rm for}&  h \geq 1
\end{array}
\ ,
\label{eq:MGS}
\end{equation}
%
%
where $I(x)$ gives the integer part of $x$, in the following sense~:
if $x=X+\delta x$ with $X$ an integer and $\delta x\in[0,1[$,
then $I(x)=X$ for $\delta x\in[0,1/2[$
and $I(x)=X+1$ for $\delta x\in[1/2,1[$.
One then trivially gets
%
%
\begin{eqnarray}
e_0(N)&=& -{1 \over 2} + \frac{2 M_0^2}{N^2} - {2h M_0 \over N} ,
\label{eq:e0_iso}\\
{2\langle S_x \rangle \over N}                            &=& 0 ,
\label{eq:sx_iso}\\
{2\langle S_y \rangle \over N}                            &=& 0 ,
\label{eq:sy_iso}\\
{2\langle S_z \rangle \over N}                            &=& {2 M_0
\over N} ,
\label{eq:sz_iso}\\
{4\left\langle S_x^2 \right\rangle \over N^2}  &=&  {4\left\langle
S_y^2 \right\rangle \over N^2} =
{2 \over N^2} \left[ {N\over 2} \left({N \over 2} +1\right) -M_0^2 \right],
\label{eq:sx2_iso}\\
{4\left\langle S_z^2 \right\rangle \over N^2}  &=& {4 M_0^2 \over N^2}.
\label{eq:sz2_iso}
\end{eqnarray}
%
%
where $e_0(N)$ is the ground-state energy per spin. The gap can also
be easily computed :
%
%
\begin{eqnarray}
\Delta(N) &=& E(N/2,M_0+s)-E(N/2,M_0) \\
              &=& {2 \over N} +2s(2M_0/N-h),
\end{eqnarray}
%
%
where $s=\mathrm{sgn}(hN/2-M_0)$ if $0<h<1$ and $s=-1$ if $h\geq 1$.
This gives $\Delta(N) = 2/N+2s(2M_0/N-h)$ in the broken phase
[which is a $2/N$-periodic function vanishing for
$h=(2p+1)/N$ and equal to $2/N$ for $h=2p/N$, $p$ denoting an
integer], whereas $\Delta(N)=2(h-1)+2/N$ in the symmetric phase.

%
%
\section{Variational and semiclassical approaches}
%
%
\label{sec:semi_classical}

In this section, we describe the semiclassical approach that can be
used to determined the phase diagram of the LMG model.\cite{Botet82,Botet83}
This analysis constitutes the zeroth order approximation and
relies on a mean-field (variational) wave function
%
%
\begin{equation}
|\psi(\theta,\phi) \rangle=\bigotimes^{N}_{l=1}
\Big[
\cos\left(\theta/2\right) e^{-\rmi \phi/2}|\! \uparrow  \rangle_{l} +
\sin\left(\theta/2\right) e^{\rmi \phi/2}|\! \downarrow \rangle_{l}
\Big],
\label{eq:ansatz}
\end{equation}
%
%
which is a coherent spin state such that
%
%
\begin{equation}
\langle {\bf S} \rangle={N \over 2} (\sin \theta \cos \phi, \sin
\theta \sin \phi, \cos \theta).
\end{equation}
%
%
In Eq.~(\ref{eq:ansatz}), kets $|\! \uparrow \rangle_{l}$ and
$|\! \downarrow \rangle_{l}$ are the eigenstates
of $\sigma_{z}^{l}$ with eigenvalues $+1$ and $-1$, respectively.
The  ground state is thus determined by minimizing the energy
%
%
\begin{equation}
  \langle H \rangle = -{(N-1)  \over 2}
  \sin^2 \theta \left(\cos^2 \phi +\gamma \sin^2 \phi \right) -h N\cos \theta.
  \label{eq:HMF}
\end{equation}
%
%
with respect to $\theta$ and $\phi$.

\begin{figure}[t]
    \centering
    \includegraphics[width=5.5cm]{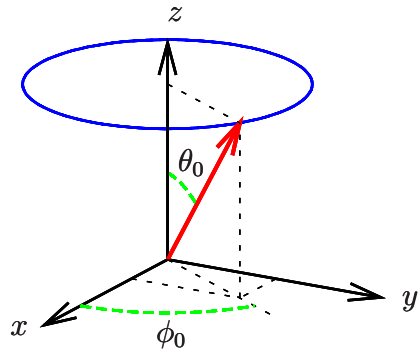}
    \caption{
      One of the infinitely degenerate mean-field ground states in the
      broken phase of the isotropic LMG model $\gamma=1$.}
    \label{fig:GS_iso_broken}
\end{figure}

\begin{figure}[t]
    \centering
    \includegraphics[width=7cm]{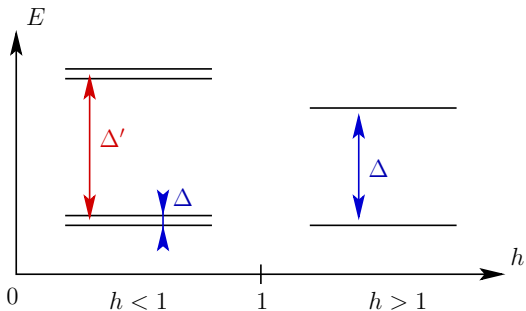}
    \caption{Schematical representation of the low-energy spectrum of the 
      anisotropic LMG model in a finite system. 
      $\Delta$ is the energy gap between the ground state and the first 
      excited state. In the broken phase $h<1$, $\Delta$ is exponentially 
      small $\Delta\sim\exp(-aN)$.\cite{Newman77}
      In this phase, $\Delta'$ is the excitation 
      energy between the two nearly degerate lowest states and the nearly 
      degenerate next two states.}
    \label{fig:spectrum}
\end{figure}

In the thermodynamical limit, this variational approach is completely
equivalent to the semiclassical
treatment  proposed by Botet and Jullien \cite{Botet83} which
consists in treating
${\bf S}$ as a classical variable.  In this limit, minimizing
$\langle H \rangle$ with respect to
$\theta$ and
$\phi$ leads to a distinction between the two phases
\begin{figure}[t]
    \centering
    \includegraphics[width=5.5cm]{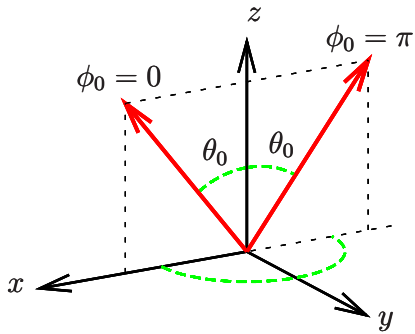}
    \caption{
      The two mean-field ground states in the broken phase of
      the anisotropic LMG model $\gamma<1$. }
    \label{fig:GS_aniso_broken}
\end{figure}
%
%
\begin{enumerate}

\item $h\geq 1$ (symmetric phase): The ground state is unique and
fully polarized in the magnetic field direction
($\theta_0=0$) for all $\gamma$, see Fig.~\ref{fig:GS_sym}.

\item $0 \leq h <1$ (broken phase):

For $\gamma \neq 1$, the ground state is twofold
degenerate ($\theta_0= \arccos h$ and $\phi_0=0$ or $\pi$) (see Fig.~\ref{fig:GS_aniso_broken}).
In the isotropic case $\gamma=1$, $\langle H \rangle$ does not depend
on $\phi$ so that the ground state is
infinitely degenerate, as shown in Fig.~\ref{fig:GS_iso_broken}.
Of course, this is due to the parametrization of the trial function.
However, if this result is clearly in contradiction with the exact
degeneracy computed in Sec.~\ref{sec:sub:isotropic} (which is unique at least for
$h \neq 0$), it points out the existence of two universality classes ($\gamma=1$ and $\gamma \neq 1$).
\end{enumerate}
%
%
%
Here, $\theta_0$ and $\phi_0$ stands for the values of $\theta$ and
$\phi$ for which $\langle H
\rangle$ is minimum.
The ground-state energy per spin is then given by:
%
%
\begin{equation}
e_0(\infty) = -\frac{1}{2}\sin^2\theta_0-h\cos\theta_0,
\label{eq:e0_mf}
\end{equation}
%
%
and the gap $\Delta$ can be evaluated  using the random phase
approximation\cite{Botet83}

%
%
\begin{equation}
\label{eq:gap_RPA}
\Delta(\infty)=\Bigg\{
\begin{array}{ccc}
2\big[(h-1)(h-\gamma)\big]^{1/2} & {\rm for}&  h \geq 1 ,\\
0                                & {\rm for}&  0\leq h < 1 .\\
\end{array}
\end{equation}
%
%
Let us note that, for the anisotropic LMG model, the result in the 
broken phase $0\leq h<1$ simply says that there are two degenerate 
ground states, and is obtained without performing a calculation. 
This gap does not correspond to the excitation 
energy around one of the two mean-field states represented in 
Fig.~\ref{fig:GS_aniso_broken}, which is finite. The latter is 
denoted as $\Delta'$ in Fig.~\ref{fig:spectrum}, whereas in the same figure, 
$\Delta$ denotes the exact gap, vanishing in the broken phase and in the 
thermodynamical limit.

The correlation functions are also easily obtained in the
thermodynamical limit since, for a
``classical" spin one simply has  $\langle S_\alpha S_\beta
\rangle=\langle S_\alpha \rangle \langle S_\beta
\rangle + O(N)$.
Note that for $\gamma=1$, one can recover the exact expressions
derived in \ref{sec:sub:isotropic} provided an
average over all possible values of $\phi_0$ is performed. For
example, one has:
%
%
\begin{eqnarray}
{4\left\langle S_x^2 \right\rangle \over N^2}  &= &
{4 \left\langle S_x \right\rangle \left\langle S_x \right\rangle\over
N^2} + O(1/N), \\
&=& \sin^2 \theta_0 \:\: \overline{\cos^2\phi_0} + O(1/N),\\
&=& {1 \over 2 } \sin^2 \theta_0 + O(1/N).
\end{eqnarray}
%
%
which coincides with Eq.~(\ref{eq:sx2_iso}) in the infinite $N$ limit.

In the thermodynamical limit, this mean-field approximation is exact
(see the comparison with the numerical
results in Sec.~\ref{sec:numerics}) and thus predicts a second-order
quantum phase transition  at the critical
field $h=1$.

For finite $N$, the exact ground state is nondegenerate and is not
given by a mean-field state
$|\psi(\theta,\phi)\rangle$ except for $h=\frac{N-1}{N} \sqrt{\gamma}$.
The twofold degeneracy (for $\gamma \neq 1$) in the broken phase
only occurs in the infinite $N$
limit where the two mean-field solutions $|\psi(\theta_0,0)\rangle$
and $|\psi(\theta_0,\pi)\rangle$
provides a (nonorthogonal) basis of the ground-state subspace.

Finally, from the entanglement point of view, $|\psi(\theta_0,\phi_0)
\rangle$ is a completely separable state
but as discussed in Sec.~\ref{sec:entanglement}, the exact
ground state has some nontrivial intrication properties (even in the
thermodynamical limit) that cannot be captured at this level. It is
thus essential to go beyond this basic approach and to compute next
orders in the $1/N$ expansion.

%

%
%
\section{First order quantum corrections}
%
%
\label{sec:firstQcorrections}

We now wish to go one step beyond the mean-field analysis of the previous
section by computing first order correction in a $1/N$ expansion. To
achieve this, we use
the Holstein-Primakoff representation of the spin operator and
truncate the resulting bosonic Hamiltonian to
lowest order. Next, we diagonalize it thanks to a Bogoliubov transformation.

%
%
\subsection{The Holstein-Primakoff representation}
%
%
\label{sec:sub:HP}

The Holstein-Primakoff representation of spin operators
\cite{Holstein40} is a useful tool to compute low-energy
corrections around a classical magnetization (see for example Ref.
\onlinecite{Auerbach94}).
The very first thing to do is thus to perform a rotation of the spin
operators around the $y$ axis, that brings
the $z$ axis along the semiclassical magnetization. This is done as follows
%
%
\begin{equation}
      \label{eq:rotation}
      \left(
        \begin{array}{c}
          S_x\\ S_y\\ S_z
        \end{array}
      \right)
      =
      \left(
        \begin{array}{ccc}
          \cos\theta_0 & 0 & \sin\theta_0\\
          0 & 1 & 0\\
          -\sin\theta_0 & 0 & \cos\theta_0
        \end{array}
      \right)
      \left(
        \begin{array}{c}
          \tS_x\\ \tS_y\\ \tS_z
        \end{array}
      \right).
\end{equation}
%
%
As explained in the previous section, $\theta_0=0$ for $h>1$ so that
${\bf S}={\bf\tS}$, and
$\theta_0=\arccos h$ for $h \leq 1$. Note that we have chosen here
the magnetization direction corresponding to
$\phi_0=0$ but the same results can be obtained by choosing $\phi_0=\pi$.

The Hamiltonian (\ref{eq:ham3}) written in terms of the ${\bf\tS}$ then reads
%
\begin{eqnarray}
      \label{eq:ham_rotated}
      H&=&\frac{1+\gamma}{2}-2hm \tS_z\nonumber\\
      &&-\frac{2}{N}\left[ (1-m^2) \tS_z^2 +\frac{m^2+\gamma}{2}
        \left({\bf \tS}^2 - \tS_z^2\right)\right]\nonumber\\
      &&+h\sqrt{1-m^2}\left( \tS_+ + \tS_-\right)\\
      &&-\frac{m\sqrt{1-m^2}}{N}\left(\tS_+ \tS_z + \tS_z \tS_+
      + \tS_- \tS_z + \tS_z\tS_-\right)\nonumber\\
      &&-\frac{m^2-\gamma}{2N}\left(\tS_+^2 + \tS_-^2\right), \nonumber
\end{eqnarray}
%
where $m={2\langle S_z \rangle/ N}= \cos \theta_0$.
The Holstein-Primakoff representation is then applied to the rotated
spin operators
%
       \begin{eqnarray}
         \tS_z&=&S-\adag a=N/2-\adag a, \label{eq:HP1}\\
         \tS_+&=&\left(2S-\adag a\right)^{1/2}a
         =N^{1/2}\left(1-\adag a/N\right)^{1/2}a, \label{eq:HP2}\\
         \tS_-&=&\adag\left(2S-\adag a\right)^{1/2}
         =N^{1/2}\adag\left(1-\adag a/N\right)^{1/2},\quad; \label{eq:HP3}
       \end{eqnarray}
%
where the bosonic creation and annihilation operators satisfy $[a,\adag]=1$.

The next step consists in inserting these expressions in Eq.~(\ref{eq:ham_rotated}),
and to expand the argument of the square roots. Keeping terms of
order $(1/N)^{-1}$, $(1/N)^{-1/2}$ and $(1/N)^{0}$ in the Hamiltonian [therefore, it is sufficient to use the
approximation $(1-\adag a/N)^{1/2}\simeq 1$] yields
%
\begin{equation}
      \label{eq:ham_order0}
      H^{(0)}=N e_0(\infty) + \delta e_0\bare + \Delta\bare \adag a
      + \Gamma\bare (\adag^2+\ana^2),
\end{equation}
%
where $e_0(\infty)$ is the mean-field ground-state energy per spin
(\ref{eq:e0_mf}),
which also reads $e_0(\infty)=(-1-2hm+m^2)/2$.
The $\bare$ superscript stands for ``bare'', and the bare couplings are
%
\begin{eqnarray}
      \label{eq:bare_couplings_ham_order0}
      \delta e_0\bare &=&\frac{1-m^2}{2},\\
      \Delta\bare &=& 2+2hm-3m^2-\gamma,\\
      \Gamma\bare &=& \frac{\gamma-m^2}{2}.
\end{eqnarray}
%

We emphasize that  the Hamiltonian contains no term proportional to
$\sqrt{N}(\adag+a)$, i.e.,  of order
$(1/N)^{-1/2}$, creating one excitation. These terms simply cancels
because the angle of the rotation
(\ref{eq:rotation}) has been chosen to bring $\tS_z$ along the
classical magnetization.
Note that this can also be achieved, as in Ref.~\onlinecite{Emary03}, for
example, by applying directly the Holstein-Primakoff mapping to the
nonrotated ${\bf S}$ operator, and performing a shift of $\adag$ and
$a$ by an appropriate constant.

%
\subsection{The Bogoliubov transformation}
%
\label{sec:sub:bogo}

The quadratic Hamiltonian (\ref{eq:ham_order0}) is easily diagonalized by
a standard  Bogoliubov transformation
%
\begin{eqnarray}
      \label{eq:bogoT}
      \adag&=&\cosh(\Theta/2) \bdag + \sinh(\Theta/2) b,\\
      a&=&\sinh(\Theta/2) \bdag + \cosh(\Theta/2) b.
\end{eqnarray}
%
where $\Theta$ is such that the Hamiltonian expressed in terms of the $b$'s does not contain a term
$\bdag^2+\anb^2$, i.e. is diagonal.
Let us set $\ve=-2\Gamma\bare/\Delta\bare$, which takes the following
values in the symmetric and broken phases
%
\begin{eqnarray}
      \label{eq:epsilon_symmetric}
      \ve(h\geq 1) &=& \frac{1-\gamma}{2h-1-\gamma},\\
      \label{eq:epsilon_broken}
      \ve(0\leq h<1) &=& \frac{h^2-\gamma}{2-h^2-\gamma}.
\end{eqnarray}
%
One easily finds that to diagonalize the Hamiltonian
(\ref{eq:ham_order0}) one must satisfy
$\tanh\Theta=\ve$. One then has
%
\begin{equation}
      \label{eq:ham_order0_diag}
      H^{(0)}=N e_0(\infty) + \delta e_0\renorm + \Delta\renorm \bdag b,
\end{equation}
%
where the $\renorm$ superscript means ``renormalized'' and the
renormalized couplings are given by
%
\begin{eqnarray}
      \label{eq:renormalized_couplings_ham_order0_e}
      \delta e_0\renorm &=&\delta e_0\bare
      +\frac{\Delta\bare}{2}(\sqrt{1-\ve^2}-1),\\
      \label{eq:renormalized_couplings_ham_order0_Delta}
      \Delta\renorm &=& \Delta\bare \sqrt{1-\ve^2}.
\end{eqnarray}
%

At this stage, it is worth noting that, in the broken phase, for
$h=\sqrt{\gamma}$, $\varepsilon$ and
thus $\Gamma\bare$ vanishes so that the Hamiltonian
(\ref{eq:ham_order0}) is readily diagonal. This
special point coincides, at this order $(1/N)^0$, with
$h=\frac{N-1}{N} \sqrt{\gamma}$ previously mentioned
(see Sec.~\ref{sec:semi_classical}) for which the product state
$|\psi(\theta_0,\phi_0) \rangle$ is the
{\it exact} ground state.

In the symmetric phase $m=1$, one thus gets
%
\begin{eqnarray}
      \label{eq:renormalized_couplings_ham_order0_sym}
      \delta e_0\renorm &=&-h+\frac{1+\gamma}{2}
      +\left[(h-1)(h-\gamma)\right]^{1/2},\\
      \Delta\renorm &=& 2\left[(h-1)(h-\gamma)\right]^{1/2}.
\end{eqnarray}
%
The last equation thus gives the gap (\ref{eq:gap_RPA}) found in the random
phase approximation.

In the broken phase $m=h$ and
%
\begin{eqnarray}
      \label{eq:renormalized_couplings_ham_order0_broken}
      \delta e_0\renorm &=&-\frac{1-\gamma}{2}
      +\left[(1-h^2)(1-\gamma)\right]^{1/2},\quad\\
      \Delta\renorm &=& 2\left[(1-h^2)(1-\gamma)\right]^{1/2}.
\end{eqnarray}
%
In this case, one must be careful about the physical interpretation of
$\Delta\renorm$. It is {\em not} the true gap of the system, opened by the
tunneling between the two classical ground states, which is
known\cite{Newman77} to be exponentially small [$\sim\exp(-aN)$]. 
It is thus not the quantity denoted as $\Delta$ in Fig.~\ref{fig:spectrum}, 
but the one denoted as $\Delta'$, which is 
the excitation energy in the vicinity of one of the classical magnetization
[up to some terms of order $\exp(-aN)$ which are not accessible in the $1/N$
expansion]. It is thus obtained by computing the energy
difference between the ground state and the second excited state. 
For the same reasons, the ground-state energy as given by 
$\delta e_0\renorm$ is only valid up to exponentially small terms.

Finally, we can compute the first quantum correction to the observables by
using Eq.~(\ref{eq:rotation}) and the expressions of the ${\bf \tilde{S}}$
operator truncated to lowest order and expressed with the $b$'s
%
\begin{eqnarray}
      \label{eq:tS_order0}
      \tS_x &\simeq& \frac{N^{1/2}}{2}\left(\adag+a\right)\nonumber\\
      &=& \frac{N^{1/2}}{2}\left(\frac{1+\ve}{1-\ve}\right)^{1/4}
      \left( \bdag+b \right),\\
      \tS_y &\simeq& \frac{\rmi N^{1/2}}{2}\left(\adag-a\right)\nonumber\\
      &=& \frac{\rmi N^{1/2}}{2}\left(\frac{1-\ve}{1+\ve}\right)^{1/4}
      \left( \bdag-b \right),\\
      \tS_z &=& \frac{N}{2}+\frac{1}{2}\left(1-\frac{1}{\sqrt{1-\ve^2}}\right)
      - \frac{1}{\sqrt{1-\ve^2}} \bdag b \nonumber\\
      &&- \frac{\ve}{2\sqrt{1-\ve^2}}\left( \bdag^2+\anb^2\right).
\end{eqnarray}
%
The one spin averages in the rotated basis deduced from these formulas read
%
\begin{eqnarray}
      \label{eq:one_spin_avg_rotated_order0_xy}
      2\langle \tS_x \rangle/N &=& 2\langle \tS_y \rangle/N = 0
      +O(1/N),\\
      \label{eq:one_spin_avg_rotated_order0_z}
      2\langle \tS_z \rangle/N &=& 1+\frac{1}{N}
      \left(1-\frac{1}{\sqrt{1-\ve^2}}\right)+O(1/N^2).\quad
\end{eqnarray}
%
In the symmetric phase, these expressions can be used by simply replacing
the $\tS$ by $S$, and by assigning the expression (\ref{eq:epsilon_symmetric})
to $\ve$.
To obtain the results in the broken phase,
one should pay attention to the fact that the corrections to
$\langle \tS_x \rangle$ and $\langle \tS_y \rangle$ on the one hand
and to $\langle \tS_z \rangle$ on the other, are not of the same order.
One then gets
%
\begin{eqnarray}
      2\langle S_x \rangle/N &=& \sqrt{1-m^2},\\
      2\langle S_y \rangle/N &=& 0,\\
      2\langle S_z \rangle/N &=& m,
\end{eqnarray}
%
which are nothing but the mean-field results.
In order to obtain the first quantum correction to these results in the
broken phase, one must also be able to compute the $1/N$ corrections to
$2\langle \tS_x \rangle/N$ and to $2\langle \tS_y \rangle/N$.
Such a calculation requires that we go beyond the Bogoliubov transformation and
will be presented in Sec.~\ref{sec:higherQcorrections}. It is nonetheless
possible to extract {\em some of} the first quantum corrections to spin-spin
correlation function in both phases
%
\begin{eqnarray}
      \label{eq:two_spin_avg_rotated_order0_xx}
      4\langle \tS_x^2 \rangle/N^2&=&\frac{1}{N}
      \left(\frac{1+\ve}{1-\ve}\right)^{1/2},\\
      \label{eq:two_spin_avg_rotated_order0_yy}
      4\langle \tS_y^2 \rangle/N^2&=&\frac{1}{N}
      \left(\frac{1-\ve}{1+\ve}\right)^{1/2},\\
      \label{eq:two_spin_avg_rotated_order0_zz}
      4\langle \tS_z^2 \rangle/N^2&=&1+\frac{2}{N}
      \left(1-\frac{1}{\sqrt{1-\ve^2}}\right),\\
      \label{eq:two_spin_avg_rotated_order0_xy}
      4\langle \tS_x \tS_y \rangle/N^2&=&4\langle \tS_y \tS_x \rangle^*/N^2
      =\frac{\rmi}{N},\\
      \label{eq:two_spin_avg_rotated_order0_xz}
      4\langle \tS_x \tS_z \rangle/N^2&=&4\langle \tS_z \tS_x \rangle^*/N^2
      =O(1/N),\\
      \label{eq:two_spin_avg_rotated_order0_yz}
      4\langle \tS_y \tS_z \rangle/N^2&=&4\langle \tS_z \tS_y \rangle^*/N^2
      =O(1/N).
\end{eqnarray}
%
Here again, the Bogoliubov transformation is a too simple calculation to
obtain all the $1/N$ corrections. In the symmetric phase however, we know
that the spin-flip symmetry is not broken, so that
Eqs.~(\ref{eq:two_spin_avg_rotated_order0_xz}) and
(\ref{eq:two_spin_avg_rotated_order0_yz}) in fact vanish.
In the broken phase, although Eq.~(\ref{eq:two_spin_avg_rotated_order0_yz}) still
vanishes, Eq.~(\ref{eq:two_spin_avg_rotated_order0_xz}) is nonvanishing.
We refer the reader to Appendix \ref{app:results} where we have gathered 
the first $1/N$ corrections (computed as explained in the next section).
Let us mention that sticking to the results
(\ref{eq:two_spin_avg_rotated_order0_xx})-
(\ref{eq:two_spin_avg_rotated_order0_yz}) presently available, the
only two-spin correlations we can compute in
both phases, and better than in the mean-field treatment is
%
\begin{equation}
      4\langle S_y^2 \rangle/N^2 = 4\langle \tS_y^2 \rangle/N^2
      = \frac{1}{N}\left(\frac{1-\ve}{1+\ve}\right)^{1/2}.
\label{eq:sy2_1}
\end{equation}
%
Luckily, as explained in Sec.~\ref{sec:sub:entanglement_th_lim}, 
this is precisely the correlation function that is
required to compute the rescaled concurrence in the parameter regime 
$h\geq 0$ for $\gamma\leq 0$ or $h\geq\sqrt{\gamma}$ for $\gamma\geq 0$ 
[see Eq.~(\ref{eq:conc2})].
However, when $\gamma\geq 0$ and $h\geq\sqrt{\gamma}$, one needs to know the 
$1/N$ corrections to $\langle S_z \rangle$ and $\langle S_z^2 \rangle$ 
[see Eq.~(\ref{eq:conc3})], which can not be computed with the Bogoliubov transform.

Of course, it would be tempting to go beyond these first-order
corrections pushing the expansion of the spin
operator to the next order. Unfortunately, from the order
$(1/N)^{1/2}$, the  Hamiltonian is no more quadratic
and, consequently,  cannot be diagonalized by a Bogoliubov
transformation. This clearly calls for a
more sophisticated method which constitutes the aim of this 
article and is presented in the next section.

%
%
\section{Higher order quantum corrections}
%
%
\label{sec:higherQcorrections}

In this section, we show how to to go beyond the first-order quantum
corrections given previously.
We explain how to compute higher-order corrections in the $1/N$
expansion thanks to the CUTs technique. These results are used, together
with a scaling argument, in order to extract
the finite-size scaling exponents of various physical quantities at
critical coupling. The analytical calculation
of these exponents is the central result of this study.

%
\subsection{Introduction to continuous unitary transformations}
%
\label{sec:sub:intro_CUTs}

To compute higher order terms in the $1/N$ expansion, we need to
diagonalize the Hamiltonian
(\ref{eq:ham_rotated}) truncated at the corresponding order, i.e. to
perform a unitary
transformation that brings the Hamiltonian to a diagonal form.
Finding such an unitary transformation
is clearly not  an easy task. It was the idea of
Wegner\cite{Wegner94} ten years ago, and independently
of G{\l}azek and Wilson\cite{Glazek93,Glazek94} to replace the single step
unitary transformation by an infinite product of infinitesimal
transformations, that is by CUTs.
In this way, one does not have to solve an algebraic problem but a set of
differential equations.

We shall now briefly recall the formalism of the CUTs, and refer the reader
to Ref.~\onlinecite{Dusuel04_2} for a pedagogical introduction and
more references on this topic.
The idea of the CUTs is to diagonalize the Hamiltonian in a continuous way
starting from the original (bare) Hamiltonian $H=H(l=0)$.
A flowing Hamiltonian is thus defined by
%
\begin{equation}
H(l)=U^\dagger(l) H(0) U(l),
\label{eq:Hl}
\end{equation}
%
where $l$ is a scaling parameter such that $H(l=\infty)$ is diagonal and
$U(l)$ is a unitary transformation, i.e.,  satisfying
$U(l)U^\dagger(l)=U^\dagger(l)U(l)=1$.
A derivation of Eq.~(\ref{eq:Hl}) with respect to $l$ yields the differential
equation (called the flow equation)
%
\begin{equation}
      \label{eq:dlH}
\pal H(l)=[\eta(l),H(l)],
\end{equation}
%
where the generator of the unitary transformation $\eta(l)$ is
%
\begin{equation}
      \eta(l) = \pal U^\dagger(l) U(l) = -U^\dagger(l) \pal U(l).
\end{equation}
%
Let us remark that the spirit is a bit different to what has been done with
the Bogoliubov transform. Indeed here, $H(\infty)$ is unitary equivalent
to $H(0)$ and is diagonal in the original basis in which $H(0)$ is
nondiagonal.
Conversely, when using the Bogoliubov transform, we expressed the
same Hamiltonian $H(0)$ in two different basis (the $a$'s and the $b$'s).

One must also simultaneously keep track of the change of basis on other
operators in which one is interested.
Denoting by $\Omega=\Omega(l=0)$ such an operator in the original
basis, one defines the flowing operator
%
\begin{equation}
\Omega(l)=U^\dagger(l) \Omega(0) U(l),
\end{equation}
%
which is subject to the same flow equation as $H$, namely,
%
\begin{equation}
\pal \Omega(l)=[\eta(l),\Omega(l)].
\end{equation}
%

This allows one to compute the expectation value of any operator $\Omega$ on
an eigenstate $|\psi\rangle$ of $H$.
Indeed, one has
%
\begin{equation}
\langle\psi|\Omega|\psi\rangle=\langle\psi|U(l=\infty)\: \Omega(l=\infty) \:
U^\dagger(l=\infty)|\psi\rangle,
\end{equation}
%
where $U^\dagger(l=\infty)|\psi\rangle$ is simply the eigenstate
of the diagonal Hamiltonian $H(l=\infty)$.

We still have to give a prescription for the anti-hermitian generator
$\eta(l)$, that has to be chosen to bring the final Hamiltonian into a
diagonal form.
Wegner proposed $\eta(l)=[H_\rmd(l),H_\mathrm{od}(l)]=[H_\rmd(l),H(l)]$,
where $H_\rmd$ and $H_\mathrm{od}$ are the diagonal and off-diagonal part
of the Hamiltonian. Other generators have been proposed, and we will use the
so-called quasiparticle conserving generator which is better adapted to our
problem. We shall now give the motivation for this choice and then derive
the flow equations.

%
\subsection{The flow equations}
%
\label{sec:sub:flow_eq}

When expressed in terms of the bosonic operators $a$ and $\adag$, the
initial Hamiltonian (\ref{eq:ham_rotated}) reads
%
\begin{equation}
     \label{eq:ham_012}
     H(0)=H_0(0)+H_1^+(0) + H_1^-(0) + H_2^+(0) + H_2^-(0),
\end{equation}
%
where $H_{1,2}^-=\left({H_{1,2}^+}\right)^\dagger$ and 0, 1, or 2 subscripts
indicate the number of created ($+$) or annihilated ($-$) excitations.
When using Wegner's generator, one would take $H_0(l)$ as
the diagonal part and the rest as the off-diagonal part so that the
generator would be
%
\begin{equation}
\eta(l)=[H_0(l),H_1^+(l) + H_1^-(l) + H_2^+(l) + H_2^-(l)].
\end{equation}
%
Such a choice of generator suffers from one
major drawback: the flowing Hamiltonian $H(l)$ does not remain band diagonal
as it is initially, but contains terms creating or annihilating any number
of excitations $H_k^\pm$, $k \geq 1$.

A possible choice of generator avoiding this problem is the so-called
quasiparticle
conserving generator first proposed by Mielke\cite{Mielke98} in the
context of finite matrices and generalized to many-body physics by
Knetter and Uhrig\cite{Knetter00}. For the problem at hand, this
generator reads
%
\begin{equation}
     \label{eq:gen_MKU}
     \eta(l)=H_1^+(l) - H_1^-(l) + H_2^+(l) - H_2^-(l),
\end{equation}
%
and coincides in the symmetric phase ($H_1^\pm=0$) with the generator
proposed by Stein\cite{Stein00}.

From a practical point of view, this generator is also more convenient than
Wegner's generator because one does not have to compute a commutator
to find the generator. In addition, the flow equations are quadratic
instead of cubic in the Hamiltonians
%
\begin{eqnarray}
     \label{eq:flow_eq_general0}
     \pal H_0(l) &=& 2\Big( \left[ H_1^+(l),H_1^-(l)\right]
       + \left[ H_2^+(l),H_2^-(l)\right] \Big),\quad\\
     \label{eq:flow_eq_general1}
     \pal H_1^+(l) &=& \left[ H_1^+(l),H_0(l)\right]
     + 2\left[ H_2^+(l),H_1^-(l)\right],\\
     \label{eq:flow_eq_general2}
     \pal H_2^+(l) &=& \left[ H_2^+(l),H_0(l)\right].
\end{eqnarray}
%
One can furthermore check, for example, that no three particle term is
generated during the flow since the sum of $[H_1^+(l),H_2^+(l)]$ and
$[H_2^+(l),H_1^+(l)]$ judiciously cancels.

\begin{figure}[t]
    \centering
    \includegraphics[width=8cm]{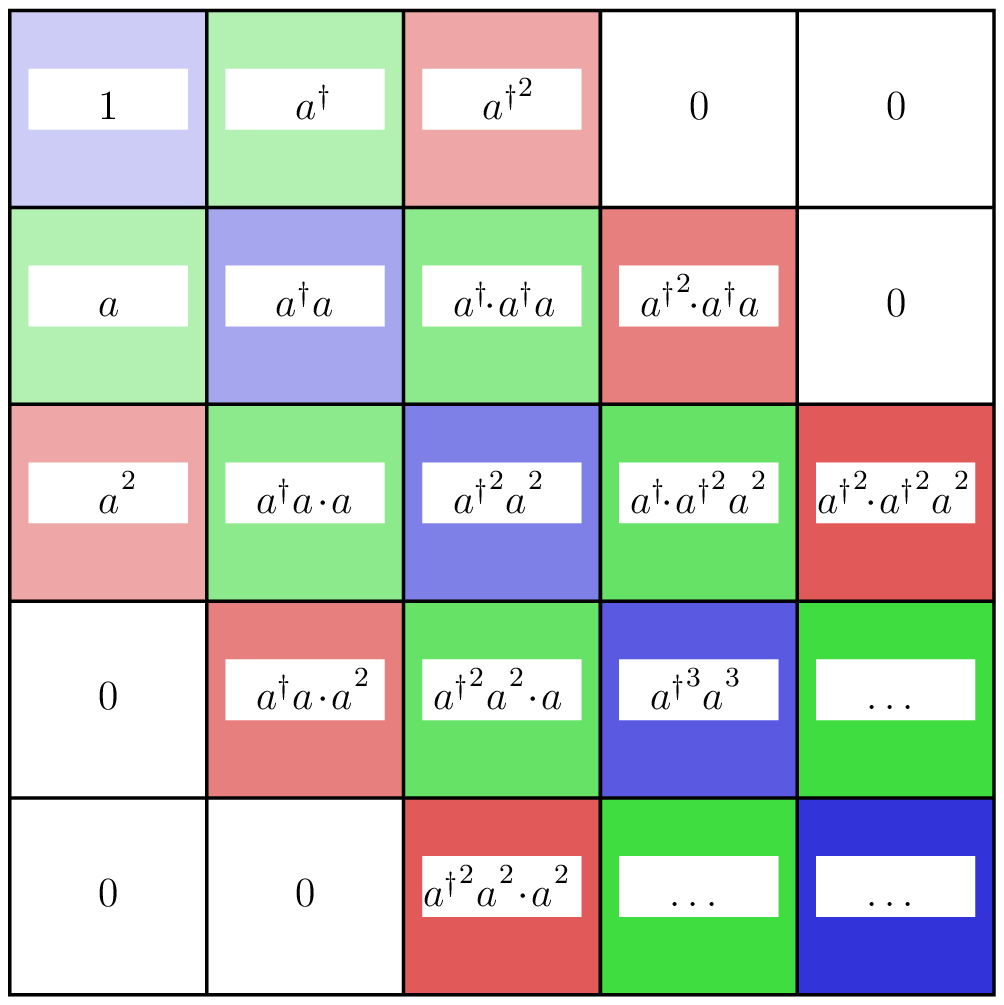}
    \caption{
      Structure of the Hamiltonian in the bosonic operators
      $a$ and $\adag$, for terms involving less than six
      $a$ or $\adag$ operators only (see the main text).}
    \label{fig:tridiag_struc}
\end{figure}
Equations (\ref{eq:flow_eq_general0})-(\ref{eq:flow_eq_general2}) are
unexploitable as they stand, because we
can only follow the flow of coupling constants, not  of operators,
and because the $1/N$ expansion is not
yet explicit. Both  points are remedied writing
%
\begin{eqnarray}
     \label{eq:H012_0}
     H_0(l) &=& \sum_{\alpha,\delta\in \nbN}
      \frac{h_{0,\alpha}^{(\delta)}(l) A_\alpha}{N^{\alpha+\delta-1}},\\
      \label{eq:H012_1}
      H_1^+(l) &=& \sum_{\alpha,\delta\in \nbN} \frac{h_{1,\alpha}^{(\delta)}(l)
        \adag A_\alpha}{N^{\alpha+\delta-1/2}},\\
      \label{eq:H012_2}
      H_2^+(l) &=& \sum_{\alpha,\delta\in \nbN} \frac{h_{2,\alpha}^{(\delta)}(l)
        \adag^2 A_\alpha}{N^{\alpha+\delta}},
\end{eqnarray}
%
where $A_\alpha=\adag^\alpha\ana^\alpha$. The couplings
$h_{k,\alpha}^{(\delta)}(l)$ are precisely the
flowing quantities we wish to follow. The $k=0, 1, 2$ subscript is
associated to the
number of created excitations $\adag^k$. The $\alpha$ subscript keeps
track of the number of $\adag$ and
$a$  operators in the $A_\alpha$ operator. Finally, the $\delta$
superscript codes
for the successive $1/N$ corrections. For example, the diagonal part of the
Hamiltonian is
%
\begin{eqnarray}
     H_0(l) &=& N h_{0,0}^{(0)}(l) + h_{0,0}^{(1)}(l)
     + \frac{1}{N} h_{0,0}^{(2)}(l) + \ldots\nonumber\\
     && + \left[ h_{0,1}^{(0)}(l) + \frac{1}{N} h_{0,1}^{(1)}(l) 
+\ldots \right ]
     \adag a\nonumber\\
     && + \left[ \frac{1}{N} h_{0,2}^{(0)}(l) + \ldots \right] \adag^2\ana^2\\
     && + \ldots,\nonumber
\end{eqnarray}
%
where only terms up to order $1/N$ have been written explicitely.

To make things even more concrete, let us write the initial conditions ($l=0$)
%
\begin{eqnarray}
     h_{0,0}^{(0)}(0) &=& \frac{-1-2hm+m^2}{2}=e_0(\infty),\\
     h_{0,0}^{(1)}(0) &=& \frac{1-m^2}{2}=\delta e_0{\bare},\\
     h_{0,0}^{(2)}(0) &=& 0,\\
     h_{0,1}^{(0)}(0) &=& 2+2hm-3m^2-\gamma=\Delta{\bare}(\infty),\\
     h_{0,1}^{(1)}(0) &=& -2+3m^2+\gamma,\\
     h_{0,2}^{(0)}(0) &=& -2+3m^2+\gamma,
\end{eqnarray}
%
and all coefficients of higher order vanish.

We have depicted the structure of the Hamiltonian
in Fig.~\ref{fig:tridiag_struc}, for all terms involving less than
six $\ana$ or $\adag$ operators. Let us emphasize that this figure 
does not represent the Hamiltonian in the eigenbasis $\{|n\rangle\}$ of 
the number operator $\adag\ana$. Let us explain on the example of the 
operator $\adag\cdot\adag^2\ana^2$ how it should be interpreted. 
This operator annihilates the states $|0\rangle$ and $|1\rangle$ 
containing zero or one excitation. The first state it does not 
annihilate is $|2\rangle$ and its action is to increase the number 
of excitations by one, and so gives a state proportional 
to $|3\rangle$. For these reasons, this operator is written 
on the first upper off-diagonal, at the position $(3,2)$. It is, however,
clear that this many-body operator also acts nontrivially on any state 
$|n\geq 2\rangle$. As another example, the zeroes shown in the figure 
mean that the Hamiltonian does not contain terms creating three or 
more excitations.

Equations (\ref{eq:H012_0})-(\ref{eq:H012_2}) can be inserted in
(\ref{eq:flow_eq_general0}-\ref{eq:flow_eq_general2}) to give the flow
equations of the various $h_{k,\alpha}^{(\delta)}(l)$ couplings
%
\begin{widetext}
     \begin{eqnarray}
       \label{eq:flow_equations_h0}
       \pal h_{0,\alpha}^{(\delta)}(l) &=& 2\sum_{n,\alpha',\delta'}
       \mathcal{C}_{\alpha',\alpha-\alpha'-1+n}^{1,1,n}
       h_{1,\alpha'}^{(\delta')}(l)
       h_{1,\alpha-\alpha'-1+n}^{(\delta-\delta'+1-n)}(l)
       +\sum_{n,\alpha',\delta'}\mathcal{C}_{\alpha',\alpha-\alpha'-2+n}^{2,2,n}
       h_{2,\alpha'}^{(\delta')}(l)
       h_{2,\alpha-\alpha'-2+n}^{(\delta-\delta'+1-n)}(l),\\
       \label{eq:flow_equations_h1}
       \pal h_{1,\alpha}^{(\delta)}(l) &=& \sum_{n,\alpha',\delta'}
       \mathcal{C}_{\alpha',\alpha-\alpha'+n}^{1,0,n}
       h_{1,\alpha'}^{(\delta')}(l)
       h_{0,\alpha-\alpha'+n}^{(\delta-\delta'+1-n)}(l) 
+2\sum_{n,\alpha',\delta'}\mathcal{C}_{\alpha',\alpha-\alpha'-1+n}^{2, 
1,n}
       h_{2,\alpha'}^{(\delta')}(l)
       h_{1,\alpha-\alpha'-1+n}^{(\delta-\delta'+1-n)}(l),\\
       \label{eq:flow_equations_h2}
       \pal h_{2,\alpha}^{(\delta)}(l) &=& \sum_{n,\alpha',\delta'}
       \mathcal{C}_{\alpha',\alpha-\alpha'+n}^{2,0,n}
h_{2,\alpha'}^{(\delta')}(l)
       h_{0,\alpha-\alpha'+n}^{(\delta-\delta'+1-n)}(l).
     \end{eqnarray}
\end{widetext}
%
The $\mathcal{C}$ coefficients arise from the computation of the commutators
%
\begin{eqnarray}
     &&\left[ \adag^{j'} A_{\alpha'} , A_{\alpha''} \ana^{j''}\right]\nonumber\\
     &=&\sum_n \mathcal{C}_{\alpha',\alpha''}^{j',j'',n}\adag^{j'-j''}
     A_{\alpha'+\alpha''+j''-n} \mbox{ if } j'\geq j'',\\
     &=&\sum_n \mathcal{C}_{\alpha',\alpha''}^{j',j'',n}
     A_{\alpha'+\alpha''+j'-n} \ana^{j''-j'} \mbox{ if } j'\leq j'',\nonumber
\end{eqnarray}
%
and are equal to
%
\begin{equation}
     \mathcal{C}_{\alpha',\alpha''}^{j',j'',n} = n! \left(
       C_{\alpha'}^n C_{\alpha''}^n
     - C_{\alpha'+j'}^n C_{\alpha''+j''}^n \right),
\end{equation}
%
$C_\alpha^n$ being the usual binomial coefficient $\alpha!/[n!(\alpha-n)!]$.
The various sums in (\ref{eq:flow_equations_h0})-(\ref{eq:flow_equations_h2})
are constrained by the fact that all subscripts and superscripts have to be
positive. Thus for example, in the first sum of Eq.~(\ref{eq:flow_equations_h0}),
$n$ runs from $0$ to $1+\delta$, $\alpha'$ from $0$ to $\alpha-1+n$, and
$\delta'$ from $0$ to $\delta+1-n$.

We shall now explain how to diagonalize the Hamiltonian order by
order in $1/N$.
The most important point is to ensure that corrections to a given
order are unchanged when going to next
orders. This means that terms of order $(1/N)^{-1}$ and
$(1/N)^{-1/2}$ must not appear in the flow
equations. Here, this is indeed the case because {\it (i)} the term
of order $(1/N)^{-1}$ is proportional to the
identity operator and {\it (ii)} the term of order $(1/N)^{-1/2}$
has been suppressed by the appropriate
choice of $\theta_0$.
Consequently, couplings of order $(1/N)^{-1}$ and $(1/N)^{-1/2}$ are simply
$h_{0,0}^{(0)}(l)=e_0(\infty)$ and $h_{1,0}^{(0)}(l)=0$ for all $l$.

At the next order $(1/N)^0$, Eqs.~(\ref{eq:flow_equations_h0})-(\ref{eq:flow_equations_h2}) give
%
\begin{eqnarray}
     \label{eq:bogot_1}
     \pal h_{0,0}^{(1)}(l) &=& -4 {h_{2,0}^{(0)}}^2(l)\\
     \pal h_{0,1}^{(0)}(l) &=& -8 {h_{2,0}^{(0)}}^2(l)\\
     \label{eq:bogot_3}
     \pal h_{2,0}^{(0)}(l) &=& -2h_{0,1}^{(0)}(l) h_{2,0}^{(0)}(l).
\end{eqnarray}
%
These equations are simply the Bogoliubov transformation, written
in a differential form \cite{Dusuel04_2}. It is a simple matter to recover
Eqs.~(\ref{eq:renormalized_couplings_ham_order0_e}) and
(\ref{eq:renormalized_couplings_ham_order0_Delta})
from the two constants of the flow,
$2h_{0,0}^{(1)}(l)-h_{0,1}^{(0)}(l)$ and
$h_{0,1}^{(0)}(l)^2-4h_{2,0}^{(0)}(l)^2$.

The same type of calculation can be performed for the observables
$\tS_x$, $\tS_y$ and $\tS_z$ as well as the correlation functions
$\tS_\alpha \tS_\beta$. As the analysis
closely resembles the one we performed for  the Hamiltonian, and as
we want the main text to be as fluid as
possible, we  have gathered the flow equations of these observables
and their derivations
in Appendix \ref{app:observables}. Let us simply mention that
contrarily to the Hamiltonian, the observables do not retain the simple
structure they exhibit in the beginning of the flow, even when using the
quasiparticle conserving generator.

The flow equations for the Hamiltonian and the observables can be integrated
{\it exactly} in both phases, at least up to the orders considered in
this work. From the solutions, one
can compute the renormalized  quantities (i.e., the values at
$l=\infty$). Note that the possibility
to get  exact $1/N$ corrections may originate from the integrability
of the LMG model and is not a generic
feature of the CUTs.
We explain how to compute the solutions to the flows of the energies and
observables in Appendix \ref{app:integration}. The results for the 
renormalized quantities are gathered in Appendix \ref{app:results} for the 
symmetric and broken phases. 
In the next subsection we explain how to make use of these results 
to compute finite-size scaling exponents.

%
\subsection{Computation of finite-size scaling exponents}
%
\label{sec:sub:fse}

\begin{figure}[t]
  \centering
  \includegraphics[width=8cm]{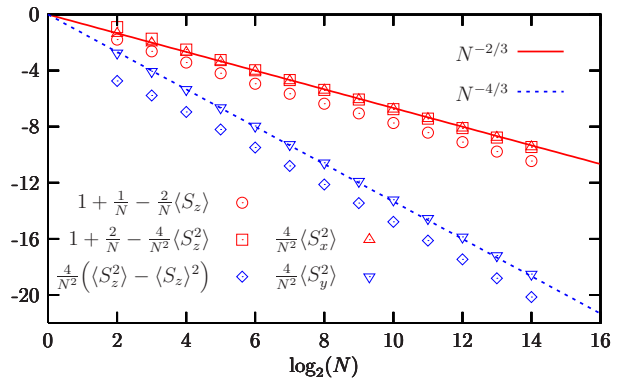}
  \caption{
    Scaling of various observables as a function of $N$
    ($\log_2-\log_2$ plot) at
    the critical point $h=1$, for $\gamma=0$..}
  \label{fig:observables}
\end{figure}

In this section, we explain how to make use of the results of the $1/N$ 
expansion to compute the finite-size exponents of the spectrum and of the 
correlation functions of the anisotropic model. 
Let us first focus on the results obtained in the symmetric phase, 
see Appendix \ref{app:sub:results_broken}. 
>From Eqs.~(\ref{eq:res_e0_sym}), (\ref{eq:res_Delta_sym}),
(\ref{eq:res_Sz_sym}), (\ref{eq:res_Sx2_sym}), (\ref{eq:res_Sy2_sym}), 
(\ref{eq:res_Sz2_sym})
one sees that any physical quantities $\Phi$ that we have computed 
can be written as 
%
\begin{equation}
  \Phi_N(h,\gamma)=
  \Phi_N^\mathrm{reg}(h,\gamma)+\Phi_N^\mathrm{sing}(h,\gamma),
\end{equation}
%
where the superscripts ``reg" and ``sing" stands for regular and singular, respectively. 
A nonsingular contribution is understood to be 
a function of $h$ which is nonsingular at $h=1$, as well as all its 
derivatives. As an example, if $\Phi=2\langle S_z \rangle/N$ 
[see Eq.~(\ref{eq:res_Sz_sym})], the regular part is $1+1/N$ 
and the remaining forms the singular part. Now let us suppose $h$ is close 
to its critical value 1. Then all terms involving the polynomial functions 
$Q_\Phi^{(i)}$ become small compared to the terms involving the 
$P_\Phi^{(i)}$'s, by a factor 
$\Xi(h,\gamma)=(h-1)(h-\gamma)$. In this limit, one can write 
%
\begin{equation}
  \Phi_N^\mathrm{sing}(h,\gamma)\simeq 
  \frac{\Xi(h,\gamma)^{\xi_\Phi}}{N^{n_\Phi}}
  \mathcal{F}_\Phi\left[N\Xi(h,\gamma)^{3/2},\gamma\right],
\end{equation}
%
where $\mathcal{F}_\Phi$ is the scaling function for the physical quantity 
$\Phi$, depending only on the scaling variable $N\Xi(h,\gamma)^{3/2}$ 
[but not separately on $N$ and $\Xi(h,\gamma)$], $\xi_\Phi$ and $n_\Phi$ 
are exponents that we list in Table~\ref{tab:exponents}.
\begin{table}[htbp]
  \centering
  \begin{tabular}{||c|c|c|c||}
    \hline
    \hline
    $\Phi$ & $\xi_\Phi$ & $n_\Phi$ & $n_\Phi+2\xi_\Phi/3$\\
    \hline
    \hline
    $e_0$ & 1/2 & 1 & 4/3\\
    \hline
    $\Delta$ & 1/2 & 0 & 1/3\\
    \hline
    $2\langle S_z \rangle/N$ & -1/2 & 1 & 2/3\\
    \hline
    $4\langle S_x^2 \rangle/N^2$ & -1/2 & 1 & 2/3\\
    \hline
    $4\langle S_y^2 \rangle/N^2$ & 1/2 & 1 & 4/3\\
    \hline
    $4\langle S_z^2 \rangle/N^2$ & -1/2 & 1 & 2/3\\
    \hline
    \hline
  \end{tabular}
  \caption{Table of exponents (see the text for their signification) 
    for the physical quantities $\Phi$ considered in this work.}
  \label{tab:exponents}
\end{table}
Now we use the fact that there can be no singularity in any physical 
quantity in a finite-size system. This implies that the singularity of 
$\Xi(h,\gamma)^{\xi_\Phi}$ has to be canceled by the one of 
$\mathcal{F}_\Phi\left[N\Xi(h,\gamma)^{3/2},\gamma\right]$. 
Thus one must have 
$\mathcal{F}_\Phi\left[x,\gamma\right]\sim x^{-2\xi_\Phi/3}$, which in turn 
implies the following finite-size scaling
\begin{equation}
  \Phi_N^\mathrm{sing}(h=1,\gamma)\sim 
  \frac{a_\Phi}{N^{n_\Phi+2\xi_\Phi/3}},
\end{equation}
where the exponents $n_\Phi+2\xi_\Phi/3$ are listed in the last column of 
Table~\ref{tab:exponents}, and the $a_\Phi$'s are constants which
cannot be determined with this scaling argument. These finite-size exponents 
are compatible with the values obtained from numerical diagonalization and 
discussed in the next section  (see Figs.~\ref{fig:gse_and_gap} 
and \ref{fig:observables}).

Let us finally mention that the exponents for the ground-state energy per 
spin $e_0$ and for the gap $\Delta$ can also be computed following the 
same line of reasoning, from the results obtained in the broken phase 
[see Eqs.~(\ref{eq:res_e0_broken}) and (\ref{eq:res_Delta_broken})]. The 
only difference is that one has to replace the variable  $\Xi(h,\gamma)=(h-1)(h-\gamma)$ by $\Psi(h)=1-h^2$. 
For the observables we did not go to sufficiently high order to perform the same analysis. However, we can
argue that from the  results for $e_0$ and $\Delta$  that the scaling variable is $N\Psi(h)^{3/2}$. Then, one
can easily show that we recover all exponents listed in Table \ref{tab:exponents}.

%
%
\section{Numerical results}
%
%
\label{sec:numerics}

In order to check the validity of the approach used here to compute
the critical exponents, we
have numerically diagonalized $H$ in the sector $S=N/2$ where the
low-energy states lie. The
dimension of this subspace is $N+1$ but the spin-flip symmetry
(\ref{eq:symmetries})
further divides by 2 the dimension of the matrices to be
diagonalized. These simplifications
allows one to easily investigate rather large system size (up to
$N=2^{20}$ spins for the energy
spectrum and $N=2^{14}$ spins for observables  requiring eigenstates).
For each system size $N$, we have computed the ground-state energy $e_0$, the energy gap $\Delta$, the
magnetization $\langle S_z \rangle$ and the correlation functions $\langle S_x^2\rangle$, $\langle S_y^2
\rangle$, and $\langle S_z^2 \rangle$ at the critical point $h=1$.
For clarity, we only present the results for $\gamma=0$ but we have
checked that the scaling exponents are the same for any $\gamma\neq 1$ as expected from our analysis, the
peculiar case $\gamma=1$ being discussed in Sec.~\ref{sec:sub:isotropic}.

\begin{figure}[t]
  \centering
  \includegraphics[width=8cm]{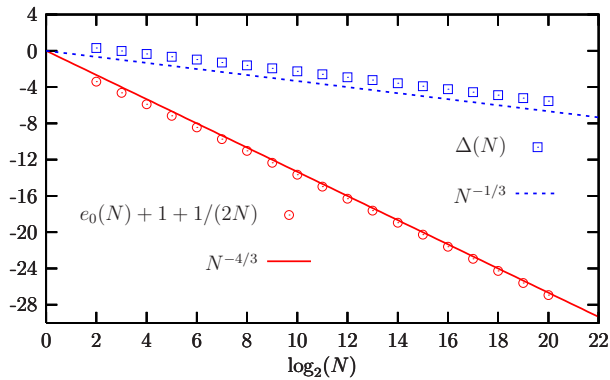}
  \caption{
    Log$_2$-$\log_2$ plot of the ground-state energy per spin $e_0$ and
    the enery gap $\Delta$ at
    the critical point $h=1$ as a function of $N$, for $\gamma=0$.}
  \label{fig:gse_and_gap}
\end{figure}

As can be seen in Fig. \ref{fig:gse_and_gap}, the leading nontrivial 
finite-size corrections for $e_0$ and $\Delta$ are indeed proportional 
to $N^{-4/3}$ and $N^{-1/3}$. 
Note that this latter result was already observed numerically 
by Botet and Jullien \cite{Botet83} twenty years ago but, 
to our knowledge, never proven by analytical arguments. 

%
%
%
%

Our results are in contradiction with a recent study by Reslen and co-workers \cite{Reslen04} who computed
numerically some correlation functions. They indeed obtained that
$2 \langle S_z^2\rangle/N-1 \sim N^{-0.55\pm 0.01}$. The source of this discrepancy comes from the too small
system size they investigated $(100 \leq N \leq 500)$. Indeed, in this range, we have checked that
the scaling exponent is indeed close to 0.55, but for larger size as those considered here ($N\sim 2^{14}$), we
found numerically that the fitted exponent gets closer and closer from 2/3 which is the value predicted by our
approach. 
%

%
%
\section{Entanglement properties}
%
%
\label{sec:entanglement}

The entanglement properties in the LMG model have recently attracted
much attention
\cite{Vidal04_1,Vidal04_2,Vidal04_3,Latorre04_3,Reslen04}. For the
ferromagnetic case considered
here, several quantities have been used to characterize the
ground-state entanglement. First, the
concurrence \cite{Wootters98} which measures the two-spin
entanglement has been shown to develop
a cusp at the critical point \cite{Vidal04_1} with a nontrivial
scaling behavior that we shall
discuss below. Secondly, the von Neumann entropy has also been computed as a
function of both the anisotropy
and the magnetic field \cite{Latorre04_3}. Its scaling at the
critical point has suggested the
possibility of an underlying conformal theory describing the LMG
model with central charges
$c=3/2$ for $\gamma=1$ and $c=1$ otherwise.  Finally, we wish to clarify the
situation about another measure proposed by Somma {\it et al.}
\cite{Somma04}, the so-called
SU(2)-purity. For a system where all spins are equivalent as in the
LMG model, this quantity
is nothing but the square of the reduced Bloch sphere radius defined by
%
\begin{equation}
r^2=\langle \sigma_x
\rangle^2+\langle \sigma_y \rangle^2+\langle \sigma_z
\rangle^2=\left({2 \langle S_z \rangle \over
N}\right)^2
\label{eq:radius}
\end{equation}
%
%
%
%
The latter relation is due to the spin-flip symmetry
(\ref{eq:symmetries}) and is only valid for the exact finite-$N$ 
ground state. It is
thus simply connected to the magnetization in the field direction so 
that {\em the
SU(2)-purity does not bring any new information  about the
entanglement of the ground state in this model}. Furthermore, this purity is
trivially computed in the
thermodynamical limit using the semiclassical description discussed in Sec.
\ref{sec:semi_classical}. Unfortunately, for the microscopic
degrees of freedom which are the
spins half, there is no entanglement in this limit since the ground
state is a pure product state
[see {\it ansatz} (\ref{eq:ansatz})]. Clearly, Somma {\it et al.} did not
understand that the richness
of the entanglement in this model was given by the finite-$N$
corrections which we shall now discuss
in the framework of the concurrence.

Introduced by Wootters \cite{Wootters98}, the concurrence which
measures the entanglement between two
spins half is defined as follows. For a given pure state $|\psi
\rangle$, we define the
density matrix $\rho=|\psi \rangle \langle \psi|$ and the two-spin
reduced density matrix
$\rho_{i,j}$ obtained by tracing out $\rho$ over all spins except
spins $i$ and $j$.
Next, we introduce the spin-flipped density matrix $\tilde
\rho_{i,j}=(\sigma_{y}\otimes\sigma_{y})\:\rho^{\ast}_{i,j}\:
(\sigma_{y}\otimes\sigma_{y})$ where
$\rho^{\ast}_{i,j}$ is the complex conjugate of $\rho_{i,j}$. The
concurrence is then defined by
%
%
\begin{equation}
C_{i,j}=\max\left\{ 0,\mu_{1}-\mu_{2}-\mu_{3}-\mu_{4}\right\},
\label{concdef}
\end{equation}
%
%
where the $\mu_{k}$'s are the square roots of the four real eigenvalues of
$\rho_{i,j} \: \tilde{\rho}_{i,j}$, and where $\mu_{i} \geq 
\mu_{i+1}$. The concurrence vanishes if the
reduced density matrix $\rho_{i,j}$ can be
decomposed into unentangled pure states whereas it reaches its
maximal values 1 for maximally entangled (two-spin) states such as
the famous Einstein-Podolsky-Rosen
state
%
%
\begin{equation}
   \displaystyle{| {\rm EPR} \rangle={1
       \over \sqrt {2}}
     \Big(|\! \uparrow \downarrow\rangle
       - |\! \downarrow \uparrow \rangle \Big)}.
\end{equation}
%
%

In the LMG model, since all spins are equivalent, the concurrence
does not explicitely depend on $i$
and $j$ and we shall omit these indexes in the following. In addition,
since the ground state lies in the
sector $S=N/2$ the concurrence can be easily expressed in terms of
the observables computed in the
previous section. Indeed, for any symmetric state, the reduced
density matrix reads \cite{Wang02}:
%
%
\begin{equation}
\rho=\left(
\begin{array}{llll}
v_{+} & x_{+}^{*} & x_{+}^{*} & u^{*} \\
x_{+} & y & y & x_{-}^{*} \\
x_{+} & y & y & x_{-}^{*} \\
u & x_{-} & x_{-} & v_{-}
\end{array}
\right)
\label{eq:rhored}
\end{equation}
%
%
with
%
%
\begin{eqnarray}
v_{\pm } &=&\frac{N^2-2N+4\langle S_z^2\rangle \pm 4 (N-1) \langle S_z\rangle%
}{4N(N-1)},  \label{eq:aaa} \\
x_{\pm } &=&\frac{(N-1)\langle S_{+}\rangle \pm \langle
[S_{+},S_z]_{+}\rangle }{2N(N-1)},  \\
y &=&\frac{N^2-4\langle S_z^2\rangle }{4N(N-1)},  \\
u &=&\frac{\langle S_{+}^2\rangle }{N(N-1)}.  \label{eq:bbb}
\end{eqnarray}
%
%
For an eigenstate of the spin-flip operator as, for instance, the
ground state of the LMG model at
finite $N$, one further has $x_{\pm}$=0, so that the concurrence is
simply given by\cite{Wang02}
%
%
\begin{equation}
C=\left\{
\begin{array}{l}
2\max (0,|u|-y) \text{ if }2y\leq\sqrt{v_{+}v_{-}}+|u|, \\
2\max (0,y-\sqrt{v_{+}v_{-}}) \text{ if } 2y \geq \sqrt{v_{+}v_{-}}+|u|.
\end{array}
\right.   \label{eq:c2}
\end{equation}
%
%


\subsection{The thermodynamical limit}
\label{sec:sub:entanglement_th_lim}

Using the expressions of the observables gathered in
Secs.~\ref{app:sub:results_sym} and
\ref{app:sub:results_broken}, one can show that for $\sqrt{\gamma}\leq h$,
one has  $2y\leq\sqrt{v_{+}v_{-}}+|u|$ so that 
the rescaled concurrence $C_\mathrm{R}=(N-1) C$ reads\cite{Dusuel04_3}
%
%
\begin{equation} \label{eq:conc1}
  C_\mathrm{R}^{\sqrt{\gamma} \leq h}
  = {2 \over N} \Big(\left|\left\langle S_x^2
      -S_y^2\right\rangle \right|
  -N^2/4 +\left\langle S_z^2 \right\rangle \Big),
\end{equation}
%
%
which can be rewritten as
%
%
\begin{equation}
    \label{eq:conc2}
    C_\mathrm{R}^{\sqrt{\gamma} \leq h}= 1-\frac{4\left\langle 
        S_y^2\right\rangle}{N},
\end{equation}
%
%
since for $|\gamma| \leq 1$, one has $\left|\left\langle S_x^2 -
S_y^2\right\rangle \right| \geq 0$.
It is worth noting that, obviously, the concurrence $C$ goes to zero
in the thermodynamic limit since the
ground state becomes a product state. However,the relevant (nontrivial) quantity for this problem 
is the rescaled concurrence $C_\mathrm{R}$. This rescaling takes
into account the fact that the entanglement is equally shared between
all spins and shall be interpreted as the connectivity of each spin.
Using expression (\ref{eq:sy2_1}) obtained by the Bogoliubov transform, 
one can then compute the asymptotic behavior of the 
concurrence in both phases. One thus has
%
%
\begin{eqnarray}
  \label{eq:conc_sym_phase}
  C_\mathrm{R}^{h\geq 1}&=& 1- \sqrt{h-1 \over h- \gamma}, \\
  \label{eq:conc_broken_phase}
  C_\mathrm{R}^{\sqrt{\gamma}\leq h \leq 1}&=& 
  1- \sqrt{1-h^2 \over 1- \gamma}.
\end{eqnarray}
%
%
These expressions are valid for any $|\gamma|<1$ and coincides, for
$\gamma=0$, with the results found by Reslen
{\it et al.} \cite{Reslen04}. They also prove that a real cusplike
singularity occurs at the critical point as
inferred from numerical results in Ref.\cite{Vidal04_1}.
Note that in the broken phase, the concurrence vanishes for
$h=\sqrt\gamma$ which is precisely the point where
the ground state is a pure product state at finite $N$  (see Sec.
\ref{sec:semi_classical}). At this special point one has
$2y = \sqrt{v_{+}v_{-}}+|u|$ and for $h\leq\sqrt\gamma$, one finds that
$2y \geq \sqrt{v_{+}v_{-}}+|u|$ so the second expression
in Eq. (\ref{eq:c2}) must be considered to compute the concurrence, namely
\begin{widetext}
%
%
\begin{equation}
  \label{eq:conc3}
  C_\mathrm{R}^{h\leq \sqrt{\gamma}}={1\over 2N} \bigg\{ N^2-4\langle S_z^2\rangle
  -\sqrt{\Big[N(N-2)+4\langle S_z^2\rangle\Big]^2 - \Big[4 (N-1)\langle
    S_z\rangle \Big]^2}
  \Bigg\}.
\end{equation}
%
%
\end{widetext}

\begin{figure}[t]
  \centering
  \includegraphics[width=8cm]{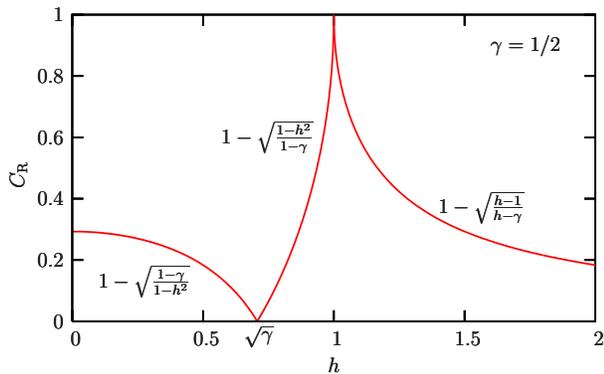}
  \caption{
    Rescaled concurrence as a function of the magnetic field, 
    in the thermodynamical limit, and for $\gamma=1/2$.}
  \label{fig:conc_th_lim}
\end{figure}

Using the expansion of $\langle S_z\rangle$
(\ref{app:sub:sub:res_sz}) and $\langle S_z^2\rangle$
(\ref{app:sub:sub:res_sz2}), one then finds:
%
%
\begin{equation}
  C_\mathrm{R}^{h\leq \sqrt{\gamma}}=1-\sqrt{1-\gamma \over 1-h^2}.
\end{equation}
%
%
We emphasize that for $h=0$, this latter expression shows that the
rescaled concurrence of the ground state is
already nontrivial since it only vanishes for $\gamma=0$. 
Furthermore, let us stress that this result is beyond the reach of the 
Bogoliubov transform. The results for the rescaled concurrence in the 
thermodynamical limit are summarized in Fig.~\ref{fig:conc_th_lim} for 
an anisotropy parameter $\gamma=1/2$.
%
%
%
%

It is important to note that, in the broken phase, we have used the expressions of the correlation functions
computed in a $1/N$  expansion around the {\it ansatz} state
$|\psi(\theta_0,\phi_0)\rangle$ [see Eq.~(\ref{eq:ansatz})] which is not the exact
ground state since it has, in particular, nonvanishing values of $\langle S_x\rangle$ and $\langle S_y\rangle$. 
The exact ground state $|\psi_0\rangle$ and the exact first excited state $|\psi_1\rangle$
indeed belong to subspaces of fixed and different spin-flip symmetry. 
However, in the broken  phase, correlation functions for these states are identical up to 
exponentially small terms $\sim\exp(-aN)$. 
Thus, for any state $|\Psi_\alpha\rangle=\cos \alpha  |\psi_0\rangle +\sin \alpha  |\psi_1\rangle$, 
the average of an operator $\Omega$ such that $\left[\Omega,\prod_i \sigma_z^i \right]=0$
(as, for instance, $S_y^2$, $S_z$ or $S_z^2$) is independent of $\alpha$ and satisfies
\begin{equation}
  \langle \Psi(\alpha) | \Omega|\Psi(\alpha)\rangle
  =\langle \psi_0 | \Omega|\psi_0\rangle
  =\langle \psi_1 | \Omega|\psi_1\rangle.
\end{equation}
This identity is valid up to exponentially small terms and  justifies our use of correlation functions computed
from the $1/N$  expansion around one of the two broken symmetry state.


\subsection{Finite-$N$ corrections}

\begin{figure}
  \centering
  \includegraphics[width=8cm]{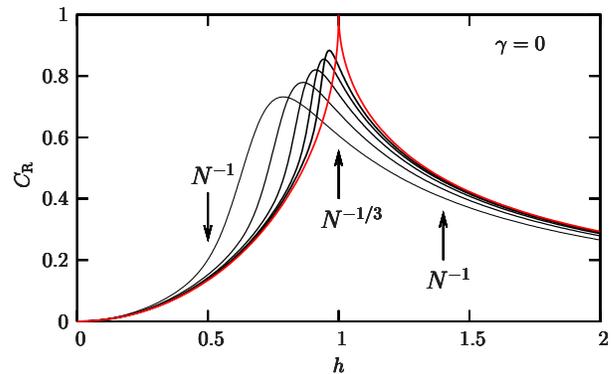}
  \caption{
    Rescaled concurrence as a function of the magnetic field, 
    for $\gamma=0$, and for system sizes $N=$16, 32, 64, 128, 256 
    and $\infty$ (bottom to top). Arrows indicate the behavior of the finite 
    size correction in various regions.}
  \label{fig:conc_vary_N}
\end{figure}

Let us now discuss the finite-$N$ corrections to the rescaled
concurrence. For $|\gamma|<1$, apart from the
critical point, the leading corrections are proportionnal to $1/N$
since the expansion of observables is not singular. {\it A contrario}, at $h=1$, using 
Eq.~(\ref{eq:conc2}) and results of Sec.~\ref{app:sub:sub:res_sy2}, one gets
%
%
\begin{equation}
C_\mathrm{R}^{h = 1 } \sim 1-a_{yy}N^{-1/3}.
\end{equation}
%
%
which is indeed the result suggested in Refs.
\cite{Vidal04_1,Reslen04}. We have summarized these scalings in Fig.
\ref{fig:conc_vary_N} for $\gamma=0$.


\subsection{The isotropic case}

For completeness, let us  briefly discuss the isotropic case. As
explained in Sec. \ref{sec:sub:isotropic}, for
$\gamma=1$, the ground state is simply the Dicke state
$\left\{N/2,M_0\rangle \right\}$ whose rescaled
concurrence is
%
%
\begin{eqnarray}
C_\mathrm{R}&=& \frac{1}{2N} \bigg\{ N^{2}-4M_0^{2}- \\
&& \sqrt{( N^{2}-4M_0^{2}) [(N-2)^{2}-4M_0^{2}]} \bigg\}  \nonumber.
\end{eqnarray}
%
%
In the thermodynamical limit, one thus has a discontinuity at the
critical point where the rescaled concurrence
jumps from 0 (for $h>1$) to 2 for ($h < 1$), and goes to 1 in the
zero field limit. Let us finally mention that this behavior can not be 
recovered by taking the limit $\gamma\to 1$ of the anisotropic results. 
Indeed, although Eq.~(\ref{eq:conc_sym_phase}) properly predicts 
$C_\mathrm{R}=0$ in the symmetric phase, the broken phase 
result (\ref{eq:conc_broken_phase}) diverges in the limit $\gamma\to 1$. 
The physical origin of this divergence is the presence of a 
Goldstone mode associate to the broken rotational symmetry along 
the $z$ axis.
%
%
%
%

%
%
\section{Conclusion}
%
%
\label{sec:conclusion}

The aim of this work was mainly twofold. We have computed the finite-size 
scaling exponents for the spectrum as well as one- and two-spin 
correlation functions of the quantum critical  LMG model. 
We have also characterized the two-spin entanglement properties of its 
ground state by computing the concurrence. 

Such studies had already been performed numerically, 
by computing the exact eigenenergies and eigenvectors. The numerical 
calculations are not too demanding since the size of the matrix to 
be diagonalized grows linearly with the number of spins $N$. Such numerical 
results were presented in Sec.~\ref{sec:numerics}, where we computed the 
eigenenergies for system sizes as large as $N=2^{20}$, and the correlation 
functions for less than $N=2^{14}$ spins. 

Our main goal however was to provide an analytical derivation of the 
simple finite-size scaling exponents observed numerically 
(Sec.~\ref{sec:sub:fse}), 
as well as formulas for the rescaled concurrence in the thermodynamical limit 
(Sec.~\ref{sec:sub:entanglement_th_lim}). 
Both of these calculations require the knowledge of $1/N$ corrections 
to the thermodynamical limit (mean-field) solution 
(Sec.~\ref{sec:semi_classical}).
On the one hand, as far as entanglement properties are concerned,  
it is clear that one has to go beyond the mean-field wave function, which 
is separable and can thus display no entanglement. 
The nontrivial entanglement properties arise 
in the finite-size system, and the physically interesting property is the 
first $1/N$ correction to the concurrence. 
On the other hand, finite-size scaling exponents at criticality can be 
extracted from the knowledge of the singularities arising in the 
$1/N$ expansion, by making use of a finite-size scaling argument. 

In Sec.~\ref{sec:firstQcorrections}, it was shown that the Bogoliubov 
transform solution of the bosonic Holstein-Primakoff representation of the 
Hamiltonian truncated to quadratic order does not capture enough of the 
$1/N$ corrections needed in this program. 
One thus has to work with a nonquadratic Hamiltonian which, in turns, 
means that one needs a method that can solve such interacting Hamiltonians. 
We have used what we think is the most efficient and elegant method 
to perform this task, namely the CUTs (Sec.~\ref{sec:sub:intro_CUTs}). 
With this method, the spectrum and correlation functions are the 
infinite time limit of some renormalization flows. 
For the LMG model, such flows can be computed exactly, yielding exact 
successive $1/N$ corrections. The integrability of the flows, which is 
not a generic feature of the CUTs, may originate in the fact that the model 
is solvable by algebraic Bethe ansatz. One should however stress that 
it is a difficult task to compute $1/N$ corrections for the spectrum from 
the Bethe ansatz solution \cite{Pan99,Links03}, and even more complicated to do it for 
correlation functions.

The combination of the $1/N$ expansion in the Holstein-Primakoff 
representation, the CUTs and a scaling argument, opens the 
possibility to compute finite-size scaling exponents as well as some 
nontrivial $1/N$ effects for other models as well. 
Of particular interest is the reduced BCS model used to describe 
superconductivity in ultrasmall grains.\cite{vonDelft01} 
This will be the object of a forthcoming publication. 
As they stand our results are also directly applicable to the Dicke model, 
which in the zero-temperature limit can be put in a one-to-one 
correspondence with the LMG model.\cite{Reslen04} 
We hope more models will turn out to be analyzable by following 
the same lines of reasoning.

\acknowledgments
We are indebted to B. Dou\c{c}ot, J. Dukelsky, S. Kirschner, J.-M. Maillard, 
D. Mouhanna, E. M\"uller-Hartmann, A. Reischl, A. Rosch and K. P. Schmidt 
for fruitful and valuable discussions.
Financial support of the DFG in SP1073 is gratefully acknowledged.

\appendix
%
%
\section{Flow equations of the observables}
%
%
\label{app:observables}

Let us denote by $\Omega(l)$ a general flowing hermitian observable.
This observable can be decomposed in
the following form:
%
%
\begin{equation}
     \Omega(l)=\Omega_0(l)+\sum_k\left[ \Omega_k^+(l) + \Omega_k^-(l) \right],
\end{equation}
%
%
where the sum runs over all nonnegative integers $k$'s  and
$\Omega_k^-={\Omega_k^+}^\dagger$. The flow equations are found by
computing $[\eta(l),\Omega(l)]$. One finds
\begin{eqnarray}
     \label{eq:flow_obs0}
     \pal \Omega_0(l) &=& \Big( \left[ H_1^+(l) , \Omega_1^-(l) \right]
       + \mbox{h.c.} \Big)\\
     && + \Big( \left[ H_2^+(l) , \Omega_2^-(l) \right]
       + \mbox{h.c.} \Big),\nonumber\\
     \label{eq:flow_obs1}
     \pal \Omega_1(l) &=& \left[ H_1^+(l), \Omega_0(l)\right]
     + {\left[ H_1^+(l), \Omega_2^-(l)\right]}^\dagger\\
     && + \left[ H_2^+(l), \Omega_1^-(l)\right]
     + {\left[ H_2^+(l), \Omega_3^-(l)\right]}^\dagger,\nonumber\\
     \label{eq:flow_obs2}
     \pal \Omega_2(l) &=& \left[ H_1^+(l), \Omega_1^+(l)\right]
     + {\left[ H_1^+(l), \Omega_3^-(l)\right]}^\dagger\\
     && + \left[ H_2^+(l), \Omega_0(l)\right]
     + {\left[ H_2^+(l), \Omega_4^-(l)\right]}^\dagger,\nonumber\\
     \label{eq:flow_obsk}
     \pal \Omega_{k\geq 3}(l) &=& \left[ H_1^+(l), \Omega_{k-1}^+(l)\right]
     + {\left[ H_1^+(l), \Omega_{k+1}^-(l)\right]}^\dagger\quad\quad\\
     && + \left[ H_2^+(l), \Omega_{k-2}^+(l)\right]
     + {\left[ H_2^+(l), \Omega_{k+2}^-(l)\right]}^\dagger.\nonumber
\end{eqnarray}

In a notation similar to Eqs.~(\ref{eq:H012_0})-(\ref{eq:H012_2}), we
have introduced
%
\begin{eqnarray}
      \Omega_0(l)&=&\sum_{\alpha,\delta\in \nbN}
      \frac{\omega_{0,\alpha}^{(\delta)}(l) A_\alpha}{N^{\alpha+\delta-1}},\\
      \Omega_k^+(l)&=&\sum_{\alpha,\delta\in \nbN}
      \frac{\omega_{k,\alpha}^{(\delta)}(l)
        \adag^k A_\alpha}{N^{\alpha+\delta+k/2-1}}.
\end{eqnarray}
%
Inserting these formulas into Eqs.~(\ref{eq:flow_obs0})-(\ref{eq:flow_obsk}) yields
\begin{widetext}
     \begin{eqnarray}
       \label{eq:flow_equations_o0}
       \pal \omega_{0,\alpha}^{(\delta)}(l) &=& 2\sum_{n,\alpha',\delta'}
       \mathcal{C}_{\alpha',\alpha-\alpha'-1+n}^{1,1,n}
       h_{1,\alpha'}^{(\delta')}(l)
       \Re\omega_{1,\alpha-\alpha'-1+n}^{(\delta-\delta'+1-n)}(l)
       + 2\sum_{n,\alpha',\delta'}
       \mathcal{C}_{\alpha',\alpha-\alpha'-2+n}^{2,2,n}
       h_{2,\alpha'}^{(\delta')}(l)
       \Re \omega_{2,\alpha-\alpha'-2+n}^{(\delta-\delta'+1-n)}(l),\quad\\
       \label{eq:flow_equations_o1}
       \pal \omega_{1,\alpha}^{(\delta)}(l) &=& \sum_{n,\alpha',\delta'}
       \mathcal{D}_{\alpha',\alpha-\alpha'+n}^{1,0,n}(l)
       h_{1,\alpha'}^{(\delta')}(l)
       \omega_{0,\alpha-\alpha'+n}^{(\delta-\delta'+1-n)}(l)
       + \sum_{n,\alpha',\delta'}
       \mathcal{C}_{\alpha',\alpha-\alpha'-1+n}^{1,2,n}
       h_{1,\alpha'}^{(\delta')}(l)
       {\omega_{2,\alpha-\alpha'-1+n}^{(\delta-\delta'+1-n)}}^*(l) \\
       && + \sum_{n,\alpha',\delta'}
       \mathcal{C}_{\alpha',\alpha-\alpha'-1+n}^{2,1,n}
       h_{2,\alpha'}^{(\delta')}(l)
       \omega_{1,\alpha-\alpha'-1+n}^{(\delta-\delta'+1-n)}(l)
       + \sum_{n,\alpha',\delta'}
       \mathcal{C}_{\alpha',\alpha-\alpha'-2+n}^{2,3,n}
       h_{2,\alpha'}^{(\delta')}(l)
       {\omega_{3,\alpha-\alpha'-2+n}^{(\delta-\delta'+1-n)}}^*(l),\nonumber\\
       \label{eq:flow_equations_ok}
       \pal \omega_{k\geq 2,\alpha}^{(\delta)}(l) &=& \sum_{n,\alpha',\delta'}
       \mathcal{D}_{\alpha',\alpha-\alpha'+n}^{1,k-1,n}(l)
       h_{1,\alpha'}^{(\delta')}(l)
       \omega_{k-1,\alpha-\alpha'+n}^{(\delta-\delta'+1-n)}(l)
       + \sum_{n,\alpha',\delta'}
       \mathcal{C}_{\alpha',\alpha-\alpha'-1+n}^{1,k+1,n}
       h_{1,\alpha'}^{(\delta')}(l)
       {\omega_{k+1,\alpha-\alpha'-1+n}^{(\delta-\delta'+1-n)}}^*(l) \\
       && + \sum_{n,\alpha',\delta'}
       \mathcal{D}_{\alpha',\alpha-\alpha'+n}^{2,k-2,n}
       h_{2,\alpha'}^{(\delta')}(l)
       \omega_{k-2,\alpha-\alpha'+n}^{(\delta-\delta'+1-n)}(l)
       + \sum_{n,\alpha',\delta'}
       \mathcal{C}_{\alpha',\alpha-\alpha'-2+n}^{2,k+2,n}
       h_{2,\alpha'}^{(\delta')}(l)
       {\omega_{k+2,\alpha-\alpha'-2+n}^{(\delta-\delta'+1-n)}}^*(l).\nonumber
     \end{eqnarray}
\end{widetext}
In the above equations, the asterix denotes complex conjugation, $\Re$ the real
part and we have defined the coefficients
%
\begin{equation}
     \mathcal{D}_{\alpha',\alpha''}^{j',j'',n} = n! \left(
       C_{\alpha'}^n C_{\alpha''+j''}^n
     - C_{\alpha'+j'}^n C_{\alpha''}^n \right),
\end{equation}
%
arising from the calculation of the commutator
%
\begin{equation}
     [\adag^{j'}A_{\alpha'},\adag^{j''}A_{\alpha''}]=
     \sum_n \mathcal{D}_{\alpha',\alpha''}^{j',j'',n}\adag^{j'+j''}
     A_{\alpha'+\alpha''-n}.
\end{equation}
%
The initial conditions for $\tS_z$, for example, are
%
\begin{equation}
\omega_{0,0}^{(0)}(l=0)=1/2, \:\:\: \omega_{0,1}^{(0)}(l=0)=-1
\end{equation}
%
and all other  coefficients are zero. Once the solution of the flow
equations for the
Hamiltonian are known, the flow equations for the observables can be
solved.

%
%
\section{Integration of the flow equations}
%
%
\label{app:integration}

In this appendix, we explain how to solve the flow equations, order by order
in the $1/N$ expansion. For the observables, we will focus on
$\tS_z$, since the computations for $\tS_x$ and $\tS_y$ differ only with
respect to the initial conditions (and also the solutions for $\tS_z$
are found to be simpler, which makes this Appendix more readable).
Furthermore, $\tS_z(l=0)$ has real coefficients, so this will remain true
for any $l$, since $\eta(l)$ also has real coefficients.
This allows us to drop the real part $\Re$ in
Eqs.~(\ref{eq:flow_equations_o0})-(\ref{eq:flow_equations_ok}).
The same applies to $\tS_x$, but not to $\tS_y$ whose
coefficients are purely imaginary.


\subsection{Order $(1/N)^{-1}$, $(1/N)^{-1/2}$, and $(1/N)^0$}

\subsubsection{Flow of the Hamiltonian}

The two first orders for the Hamiltonian have already been discussed
in the main text:
$h_{0,0}^{(0)}(l)=e_0(\infty)$ to order $(1/N)^{-1}$ and
$h_{1,0}^{(0)}(l)=0$ to order $(1/N)^{-1/2}$.
Further, we recall how crucial it is to ensure that
$h_{1,0}^{(0)}(l=0)=0$, so that this remains
true at any $l$. In the following, we will thus drop all
contributions to the flow
equations involving this term.

To order $(1/N)^0$, the flow equations are given by
Eqs.~(\ref{eq:bogot_1})-(\ref{eq:bogot_3}).
The equations for $h_{0,1}^{(0)}(l)$ and $h_{2,0}^{(0)}(l)$ are easily solved
by noticing that $h_{0,1}^{(0)}(l)^2-4h_{2,0}^{(0)}(l)^2$ is a constant of
the flow. One gets the hyperbolic solutions
%
\begin{eqnarray}
     \label{eq:sol_bogo1}
     h_{0,1}^{(0)}(l) &=&
     \frac{\Delta\renorm}{\tanh\left[ 2\Delta\renorm (l+l_0)\right]}
     =\Delta\bare f_1(l),\\
     \label{eq:sol_bogo2}
     h_{2,0}^{(0)}(l) &=&
     \frac{-\sgn(\ve)\Delta\renorm}{2\sinh\left[ 2\Delta\renorm (l+l_0)\right]}
     =\Gamma\bare f_2(l),
\end{eqnarray}
%
where $l_0$ is such that the initial conditions are fulfilled, namely
$\Delta\bare=\Delta\renorm/\tanh\left[ 2\Delta\renorm l_0\right]$.
The quantities
$\Delta\bare$, $\Gamma\bare$, $\Delta\renorm$, and $\ve$ are defined
in Sec.~\ref{sec:firstQcorrections}. The functions $f_1(l)$ and $f_2(l)$ are
defined by the two equations above, and satisfy $f_1(0)=f_2(0)=1$. Finally,
$h_{0,0}^{(1)}(l)$ is found thanks to the second constant of the flow, namely
$2h_{0,0}^{(1)}(l)-h_{0,1}^{(0)}(l)$.

There is also another useful  representation of the solutions
(\ref{eq:sol_bogo1}) and
(\ref{eq:sol_bogo2}). Let us introduce a new ``time'' scale
%
\begin{equation}
     t=\sgn(\ve)\exp\left[2\Delta\renorm (l+l_0)\right],
\end{equation}
%
with initial conditions now given at
%
\begin{equation}
     t_0=\sgn(\ve)\exp\left(2\Delta\renorm l_0\right).
\end{equation}
%
After some algebra $t_0$ can also be shown to be equal to
    %
\begin{equation}
     \label{eq:t0}
     t_0=\frac{1}{\ve}\left( 1+\sqrt{1-\ve^2}\right).
\end{equation}
%
Equations ({\ref{eq:sol_bogo1}) and (\ref{eq:sol_bogo2}) now read
%
\begin{eqnarray}
     \label{eq:sol_bogo_t1}
     h_{0,1}^{(0)}(t) &=& \Delta\renorm\frac{t^2+1}{t^2-1},\\
     \label{eq:sol_bogo_t2}
     h_{2,0}^{(0)}(t) &=& -\Delta\renorm\frac{t}{t^2-1}.
\end{eqnarray}
%
The renormalized values at $l\to\infty$ are now found by taking the limit
$t\to t_\infty=\sgn(\ve)\infty$. Let us remark that the off-diagonal coupling
$h_{2,0}^{(0)}(t)$ goes to zero and behaves as $t^{-1}$ for
$t\to t_\infty$. This will be true for all off-diagonal couplings
creating two excitations since the energy cost of such excitations, in the
thermodynamic limit and for large $t$, is nothing but $2\Delta\renorm$,
so that the couplings must vanish as $\exp(-2\Delta\renorm l)$.


\subsubsection{Flow of the $\tS_z$ observable}

To order $(1/N)^{-1}$ the flow is
%
\begin{equation}
     \pal \omega_{0,0}^{(0)}(l) = 0,
\end{equation}
%
As a consequence, $\omega_{0,0}^{(0)}(l) = \omega_{0,0}^{(0)}(0) = 1/2$.

To order $(1/N)^{-1/2}$, there is only one flow equation, namely
%
\begin{equation}
     \pal \omega_{1,0}^{(0)}(l) = -2 h_{2,0}^{(0)}(l) \omega_{1,0}^{(0)}(l),
\end{equation}
%
Since for $\tS_z$ the initial condition is $\omega_{1,0}^{(0)}(0)=0$, one
gets $\omega_{1,0}^{(0)}(l)=0$.

Working to order $(1/N)^0$ one has three flow equations to solve, which are
%
\begin{eqnarray}
     \label{eq:omega_N0_1}
     \pal \omega_{0,0}^{(1)}(l) &=& -2 h_{1,0}^{(1)} \omega_{1,0}^{(0)}(l)
     -4 h_{2,0}^{(0)}(l) \omega_{2,0}^{(0)}(l),\\
     \label{eq:omega_N0_2}
     \pal \omega_{0,1}^{(0)}(l) &=& -4 h_{1,1}^{(0)} \omega_{1,0}^{(0)}(l)
     -8 h_{2,0}^{(0)}(l) \omega_{2,0}^{(0)}(l),\\
     \label{eq:omega_N0_3}
     \pal \omega_{2,0}^{(0)}(l) &=& h_{1,0}^{(1)} \omega_{1,0}^{(0)}(l)
     -2 h_{2,0}^{(0)}(l) \omega_{0,1}^{(0)}(l).
\end{eqnarray}
%
For the $\tS_z$ observable considered here for illustration, the
previous order solution
tells us that the first terms in the three right-hand sides (RHSs) of the above
equations vanish, so that the equations are homogeneous.
Let us stress that it is not the case for $\tS_x$.
One first has to solve the subsystem (\ref{eq:omega_N0_2}) and
(\ref{eq:omega_N0_3}), and then the RHS of (\ref{eq:omega_N0_1}) is known
so that it can also be integrated. In the $t$ parametrization, one has
%
\begin{eqnarray}
     \label{eq:omega_N0_t_1}
     \pat \omega_{0,0}^{(1)}(t) &=& -\frac{2}{1-t^2} \omega_{2,0}^{(0)}(t),\\
     \label{eq:omega_N0_t_2}
     \pat \omega_{0,1}^{(0)}(t) &=& -\frac{4}{1-t^2} \omega_{2,0}^{(0)}(t),\\
     \label{eq:omega_N0_t_3}
     \pat \omega_{2,0}^{(0)}(t) &=& -\frac{1}{1-t^2} \omega_{0,1}^{(0)}(t),
\end{eqnarray}
%
which is integrated in
%
\begin{eqnarray}
     \omega_{0,0}^{(1)}(t) &=& -\frac{(t-t_0)^2}
     {(t^2-1)(t_0^2-1)},\\
     \omega_{0,1}^{(0)}(t) &=& -\frac{1 + t_0^2 - 4 t_0 t + t^2 + t_0^2 t^2}
     {(t^2-1)(t_0^2-1)},\\
     \omega_{2,0}^{(0)}(t) &=& \frac{(t-t_0)(1-t_0 t)}
     {(t^2-1)(t_0^2-1)},
\end{eqnarray}
%
satisfying the initial conditions $\omega_{0,0}^{(1)}(t_0)=0$,
$\omega_{0,1}^{(0)}(t_0)=-1$ and $\omega_{2,0}^{(0)}(t_0)=0$. From these, one
can deduce the renormalized values
%
\begin{eqnarray}
     \label{eq:omega_N0_tinf_1}
     \omega_{0,0}^{(1)}(t_\infty) &=& -\frac{1}{t_0^2-1}
     =\frac{1}{2}\left( 1-\frac{1}{\sqrt{1-\ve^2}}\right), \\
     \label{eq:omega_N0_tinf_2}
     \omega_{0,1}^{(0)}(t_\infty) &=& -\frac{t_0^2+1}{t_0^2-1}
     =-\frac{1}{\sqrt{1-\ve^2}},\\
     \label{eq:omega_N0_tinf_3}
     \omega_{2,0}^{(0)}(t_\infty) &=& -\frac{t_0}{t_0^2-1}
     =-\frac{\ve}{2\sqrt{1-\ve^2}}.
\end{eqnarray}
%
Up to a factor $1/2$, Eq.~(\ref{eq:omega_N0_tinf_1}) is nothing but the $1/N$
term in Eq.~(\ref{eq:one_spin_avg_rotated_order0_z}).


\subsection{Flow of the Hamiltonian at order $(1/N)^{1/2}$}

The flow equations read
%
\begin{eqnarray}
     \pal h_{1,0}^{(1)}(l) &=& -\left[h_{0,1}^{(0)}(l)
       + 4 h_{2,0}^{(0)}(l)\right] h_{1,0}^{(1)}(l)\nonumber\\
     && - 4 h_{2,0}^{(0)}(l) h_{1,1}^{(0)}(l),\\
     \pal h_{1,1}^{(0)}(l) &=& -\left[h_{0,1}^{(0)}(l)
       + 8 h_{2,0}^{(0)}(l)\right] h_{1,1}^{(0)}(l),
\end{eqnarray}
%
with $h_{1,1}^{(0)}(0)=2h_{1,0}^{(1)}(0)=2m\sqrt{1-m^2}$.
The second equation is independent of the first one so it can be solved
separately. Then its solution can be inserted in the first equation. This
first equation is then solved in two steps, the first being to solve the
homogeneous equation, and the second being to find a particular solution of
the inhomogeneous solution.
Again, the simplest way to solve these equations is to use the $t$ timescale.
We simply give the solutions, which read
%
\begin{widetext}
     \begin{eqnarray}
       h_{1,0}^{(1)}(t) &=& -m\sqrt{1-m^2}
       \frac{t^{1/2}(t-1)^{1/2}(t_0+1)^{3/2}(1+3t_0-3t-t_0t)}
       {t_0^{1/2}(t_0-1)^{3/2}(t+1)^{5/2}},\\
       h_{1,1}^{(0)}(t) &=& 2m\sqrt{1-m^2}\frac{t^{1/2}(t-1)^{3/2}(t_0+1)^{5/2}}
       {t_0^{1/2}(t_0-1)^{3/2}(t+1)^{5/2}}.\quad
     \end{eqnarray}
\end{widetext}
%
We see that both these off-diagonal couplings go to zero and
behave as $t^{-1/2}$ for $t\to t_\infty$.
This is a general feature of all off-diagonal couplings creating one
excitation since their energy cost, in the thermodynamic limit and for
large $t$, is nothing but $\Delta\renorm$, meaning that these couplings
have to go to zero as $\exp(-\Delta\renorm l)$.


\subsection{Next orders}

>From order $(1/N)^1$ for the Hamiltonian and $(1/N)^{1/2}$ for the
observables, things really become more involved, and it does not make
any sense any more to give the solutions in this Appendix.
However, we will explain how one can find these solutions.

Stein gave some explicit solutions for the flow
of the Hamiltonian in the symmetric phase and in the $l$ time-scale
language [up to order $(1/N)^1$] \cite{Stein00}. The solutions he
gives are found to be
polynomials in three functions: $f_1(l)$, $f_2(l)$ [see Eqs.~(\ref{eq:sol_bogo1})
and (\ref{eq:sol_bogo2})] and $f_3(l)=l+1$
[satisfying the same initial condition as $f_1$ and $f_2$,
namely, $f_3(0)=1$]. The degree of the polynom in $f_1$ can
be restricted to be one, since one has $f_1^2(l)-\ve^2 f_2^2(l)=1-\ve^2$.
In Stein's solutions, the exponential decrease of all off-diagonal couplings
creating two excitations is found to mean that one can factor out a term
$f_2(l)$.

We have checked that to the maximal orders at which we worked, an ansatz of
the following form:
%
\begin{eqnarray}
     h_{0,\alpha}^{(\delta)}(l) &=&
     P_{0,\alpha}^{(\delta)}[f_1(l),f_2(l),f_3(l)],\\
     h_{1,\alpha}^{(\delta)}(l) &=&
     f_2^{1/2}(l)P_{1,\alpha}^{(\delta)}[f_1(l),f_2(l),f_3(l)],\\
     h_{2,\alpha}^{(\delta)}(l) &=&
     f_2(l) P_{2,\alpha}^{(\delta)}[f_1(l),f_2(l),f_3(l)],
\end{eqnarray}
%
is always a solution provided the degrees in $f_2$ and $f_3$ of the
polynoms $P_0$, $P_1$,
and $P_2$ are large enough. The prefactor $f_2^{1/2}(l)$ in the solution
for $h_1$ ensures the exponential decrease of these terms creating one
excitation, with a time-scale being half the one for terms creating two
excitations.

In the $t$ time-scale language, $f_2(l)$ reads
%
\begin{equation}
     f_2(l)=\frac{t(t_0^2-1)}{t_0(t^2-1)}.
\end{equation}
%
The $P$ polynomial functions translate in fractions having the following
properties. The denominator is of the form
$(t-1)^{n_-}(t-1)^{n_+}$, with $n_-$ and $n_+$ being integers. The numerator
is a polynomial function in $t$ and in $\ln(t/t_0)$ (this logarithm is $f_3$
up to multiplicative constant).

\begin{widetext}
%
%
\section{Coefficients of the $1/N$ expansion}
%
%
\label{app:results}


In this appendix, we give the first terms of the $1/N$ expansion for
various quantities. Note that we could have only given
$\langle H \rangle/N=e_0(N)$, $\langle S_z \rangle$, and
$\langle S_z^2 \rangle$ which are sufficient to compute all
these quantities (except the gap). Indeed, one has
%
\begin{equation}
   \langle H \rangle=-\frac{2 \lambda}{N}
   \Big(\langle S_x^2 \rangle +\gamma \langle S_y^2 \rangle \Big)
   -2 h \langle S_z \rangle + {\lambda \over 2} (1+\gamma)
\end{equation}
%
and
%
\begin{equation}
   \langle{\bf S}^2 \rangle=\langle S_x^2 \rangle
   +\langle S_y^2 \rangle+\langle S_z^2 \rangle
   ={N\over 2}\left( {N\over 2}+1\right).
\end{equation}
%
However, we have considered that explicit forms of
$\langle S_x^2 \rangle$, $\langle S_y^2 \rangle$ are of interest
because of subtle cancellations of some coefficients.

For the special case $\gamma=-1$, we emphasize that CUTs have already been used by several groups \cite{Pirner98,Scholtz03,Kriel05}, notably to compute the $1/N$ corrections \cite{Stein00}. 
In addition, these corrections have also been obtained by a perturbative method \cite{Dzhioev04}.

\subsection{Symmetric phase}
\label{app:sub:results_sym}

In what follows, we denote by $\Xi(h,\gamma)=(h-1)(h-\gamma)$.


\subsubsection{Ground-state energy per spin $e_0(N)$}

%
\begin{eqnarray}
  \label{eq:res_e0_sym}
    e_0(N)&=&-h+\frac{1}{N}\left[\Xi(h,\gamma)^{1/2}
      +\frac{1}{2}(1+\gamma)-h\right]+\frac{1}{N^2}\left[
      \frac{P_e^{(1)}(h,\gamma)}{\Xi(h,\gamma)}+\frac{Q_e^{(1)}(h,
        \gamma)}{\Xi(h,\gamma)^{1/2}}\right]\\
    &&+(1-\gamma)^2\Bigg\{\frac{1}{N^3}\left[
      \frac{P_e^{(2)}(h,\gamma)}{\Xi(h,\gamma)^{5/2}}+\frac{Q_e^{(2)}(h,
        \gamma)}{\Xi(h,\gamma)^{2}}\right]
    +\frac{1}{N^4}\left[
      \frac{P_e^{(3)}(h,\gamma)}{\Xi(h,\gamma)^{4}}
      +\frac{Q_e^{(3)}(h,\gamma)}{\Xi(h,\gamma)^{7/2}}\right]\Bigg\}+
    O\left(\frac{1}{N^5}\right),\nonumber
\end{eqnarray}
%
where the $P$'s and $Q$'s are polynomial functions of $h$ and
$\gamma$ that we list below
%
\begin{eqnarray}
    P_e^{(1)}(h,\gamma)&=&\frac{1}{2}(1+\gamma)(h^2+\gamma)
    -\frac{1}{8}h(1+14\gamma+\gamma^2),\\
    Q_e^{(1)}(h,\gamma)&=&-\frac{1}{2}(1+\gamma)h+\gamma,\\
    P_e^{(2)}(h,\gamma)&=&\frac{1}{16}(1+\gamma)h(h^2-\gamma)
    -\frac{1}{64}\Big[ 24h^4+(1-42\gamma+\gamma^2)h^2+16\gamma^2\Big],\\
    Q_e^{(2)}(h,\gamma)&=&\frac{3}{8}h(h^2-\gamma),\\
    P_e^{(3)}(h,\gamma)&=&\frac{3}{128}(1+\gamma)\Big[32h^6
    +(1+14\gamma+\gamma^2)h^4+\gamma(1-66\gamma+\gamma^2)h^2
    +16\gamma^3\Big]\\
    &&-\frac{3}{512}h\Big[ 20(1+30\gamma+\gamma^2)h^4
    +(1+36\gamma-842\gamma^2+36\gamma^3+\gamma^4)h^2
    -4\gamma^2(11-54\gamma+11\gamma^2)\Big],\nonumber\\
    Q_e^{(3)}(h,\gamma)&=&-\frac{1}{128}(1+\gamma)h\Big[96h^4
    +(1-34\gamma+\gamma^2)h^2-64\gamma^2\Big]\\
    &&-\frac{1}{64}\Big[(4-184\gamma+4\gamma^2)h^4
    -\gamma(5-202\gamma+5\gamma^2)h^2-16\gamma^3\Big].\nonumber
\end{eqnarray}
%


\subsubsection{Gap $\Delta(N)$}

%
\begin{eqnarray}
  \label{eq:res_Delta_sym}
    \Delta(N)&=&2\Xi(h,\gamma)^{1/2}+\frac{1}{N}\left[
      \frac{P_\Delta^{(1)}(h,\gamma)}{\Xi(h,\gamma)}+\frac{Q_\Delta^{(1)}(h,
        \gamma)}{\Xi(h,\gamma)^{1/2}}\right]\\
    &&+(1-\gamma)^2\Bigg\{\frac{1}{N^2}\left[
      \frac{P_\Delta^{(2)}(h,\gamma)}{\Xi(h,\gamma)^{5/2}}
      +\frac{Q_\Delta^{(2)}(h,
        \gamma)}{\Xi(h,\gamma)^{2}}\right]+\frac{1}{N^3}\left[
      \frac{P_\Delta^{(3)}(h,\gamma)}{\Xi(h,\gamma)^{4}}
      +\frac{Q_\Delta^{(3)}(h,\gamma)}{\Xi(h,\gamma)^{7/2}}\right]\Bigg\}+
    O\left(\frac{1}{N^4}\right),\nonumber
\end{eqnarray}
%
with
%
\begin{eqnarray}
    P_\Delta^{(1)}(h,\gamma)&=&2(1+\gamma)(h^2+\gamma)
    -\frac{1}{2}h(1+14\gamma+\gamma^2),\\
    Q_\Delta^{(1)}(h,\gamma)&=&-(1+\gamma)h+2\gamma,\\
    P_\Delta^{(2)}(h,\gamma)&=&\frac{1}{8}(1+\gamma)h(h^2-\gamma)
    -\frac{1}{8}\Big[ 18h^4+(1-36\gamma+\gamma^2)h^2+16\gamma^2\Big],\\
    Q_\Delta^{(2)}(h,\gamma)&=&\frac{3}{2}h(h^2-\gamma),\\
    P_\Delta^{(3)}(h,\gamma)&=&\frac{3}{64}(1+\gamma)\Big[144h^6
    +(7-62\gamma+7\gamma^2)h^4+\gamma(7-222\gamma+7\gamma^2)h^2
    +112\gamma^3\Big]\\
    &&-\frac{3}{256}h\Big[ 2560\gamma h^4
    +(7+212\gamma-4534\gamma^2+212\gamma^3+7\gamma^4)h^2
    -128\gamma^2(1-14\gamma+\gamma^2)\Big],\nonumber\\
    Q_\Delta^{(3)}(h,\gamma)&=&-\frac{1}{16}(1+\gamma)h\Big[84h^4
    +(1-34\gamma+\gamma^2)h^2-52\gamma^2\Big]\\
    &&-\frac{1}{16}\Big[(5-314\gamma+5\gamma^2)h^4
    -7\gamma(1-50\gamma+\gamma^2)h^2-32\gamma^3\Big].\nonumber
\end{eqnarray}
%


\subsubsection{One-spin expectation value $\langle S_z \rangle$}
\label{app:sub:sub:res_sz}

%
\begin{eqnarray}
  \label{eq:res_Sz_sym}
    \frac{2\langle S_z\rangle}{N}&=&1+\frac{1}{N}\left[
      \frac{P_z^{(1)}(h,\gamma)}{\Xi(h,\gamma)^{1/2}}+1\right]\\
    &&+(1-\gamma)^2\Bigg\{\frac{1}{N^2}\left[
      \frac{P_z^{(2)}(h,\gamma)}{\Xi(h,\gamma)^{2}}
      +\frac{Q_z^{(2)}(h,
        \gamma)}{\Xi(h,\gamma)^{3/2}}\right]+\frac{1}{N^3}\left[
      \frac{P_z^{(3)}(h,\gamma)}{\Xi(h,\gamma)^{7/2}}
      +\frac{Q_z^{(3)}(h,\gamma)}{\Xi(h,\gamma)^{3}}\right]\Bigg\}
    +O\left(\frac{1}{N^4}\right),\nonumber
\end{eqnarray}
%
with
%
\begin{eqnarray}
    P_z^{(1)}(h,\gamma)&=&\frac{1}{2}(1+\gamma)-h,\\
    P_z^{(2)}(h,\gamma)&=&\frac{3}{8}(h^2-\gamma),\\
    Q_z^{(2)}(h,\gamma)&=&-\frac{1}{4}h,\\
    P_z^{(3)}(h,\gamma)&=&-\frac{1}{128}(1+\gamma)\Big[56h^4
    -(1-98\gamma+\gamma^2)h^2-88\gamma^2\Big]\\
    &&-\frac{1}{64}h\Big[24h^4+(1-226\gamma+\gamma^2)h^2
    -8\gamma(1-19\gamma+\gamma^2)\Big],\nonumber\\
    Q_z^{(3)}(h,\gamma)&=&\frac{3}{8}(1+\gamma)h(h^2+\gamma)
    +\frac{3}{8}(h^4-6\gamma h^2+\gamma^2\Big).
\end{eqnarray}
%


\subsubsection{Two-spin expectation value $\langle S_x^2 \rangle$}
\label{app:sub:sub:res_sx2}

%
\begin{eqnarray}
  \label{eq:res_Sx2_sym}
    \frac{4\langle S_x^2\rangle}{N^2}&=&(h-\gamma)\Bigg\{
      \frac{1}{N} \frac{1}{\Xi(h,\gamma)^{1/2}}
    +\frac{1}{N^2}\left[
      \frac{P_{xx}^{(2)}(h,\gamma)}{\Xi(h,\gamma)^{2}}
      +\frac{Q_{xx}^{(2)}(h,
        \gamma)}{\Xi(h,\gamma)^{3/2}}\right]\\
    &&+\frac{1}{N^3}\left[
      \frac{P_{xx}^{(3)}(h,\gamma)}{\Xi(h,\gamma)^{7/2}}
      +\frac{Q_{xx}^{(3)}(h,
        \gamma)}{\Xi(h,\gamma)^{3}}\right]\Bigg\}
    +O\left(\frac{1}{N^4}\right)\nonumber,
\end{eqnarray}
%
with
%
\begin{eqnarray}
    P_{xx}^{(2)}(h,\gamma)&=&-\frac{1}{4}\Big[4h^3-2(1+5\gamma)h^2
    +(1+8\gamma+3\gamma^2)h-4\gamma\Big],\\
    Q_{xx}^{(2)}(h,\gamma)&=&\frac{1}{2}\Big[2h^2-(1+3\gamma)h
    +2\gamma\Big],\\
    P_{xx}^{(3)}(h,\gamma)&=&\frac{1}{64}\Big[ 96(1-\gamma)h^5
    -128\gamma(1-\gamma)h^4+4(3-65\gamma+89\gamma^2-27\gamma^3)h^3\\
    &&-(3-76\gamma-62\gamma^2+148\gamma^3-13\gamma^4)h^2
    -4\gamma(3-33\gamma+49\gamma^2-19\gamma^2)h
    -16\gamma^2(3-2\gamma-\gamma^2)\Big],\nonumber\\
    Q_{xx}^{(3)}(h,\gamma)&=&-\frac{3}{4}(1-\gamma)h\Big[ 2h^3
    -2\gamma h^2-\gamma(3-\gamma)h+\gamma(1+\gamma)\Big].
\end{eqnarray}
%


\subsubsection{Two-spin expectation value $\langle S_y^2 \rangle$}
\label{app:sub:sub:res_sy2}

%
\begin{eqnarray}
  \label{eq:res_Sy2_sym}
    \frac{4\langle S_y^2\rangle}{N^2}&=&\frac{1}{h-\gamma}\Bigg\{
    \frac{1}{N}\Xi(h,\gamma)^{1/2}+\frac{1}{N^2}\left[
      \frac{P_{yy}^{(2)}(h,\gamma)}{\Xi(h,\gamma)}
      +\frac{Q_{yy}^{(2)}(h,
        \gamma)}{\Xi(h,\gamma)^{1/2}}\right]\\
    &&+\frac{1}{N^3}\left[
      \frac{P_{yy}^{(3)}(h,\gamma)}{\Xi(h,\gamma)^{5/2}}
      +\frac{Q_{yy}^{(3)}(h,
        \gamma)}{\Xi(h,\gamma)^{2}}\right]\Bigg\}
    +O\left(\frac{1}{N^4}\right)\nonumber,
\end{eqnarray}
%
with
%
\begin{eqnarray}
    P_{yy}^{(2)}(h,\gamma)&=&-\frac{1}{4}\Big[4h^3-2(5+\gamma)h^2
    +(3+8\gamma+\gamma^2)h-4\gamma^2\Big],\\
    Q_{yy}^{(2)}(h,\gamma)&=&\frac{1}{2}\Big[2h^2-(3+\gamma)h
    +2\gamma\Big],\\
    P_{yy}^{(3)}(h,\gamma)&=&-\frac{1}{64}\Big[ 96(1-\gamma)h^5
    -128(1-\gamma)h^4+4(27-89\gamma+65\gamma^2-3\gamma^3)h^3\\
    &&-(13-148\gamma+62\gamma^2+76\gamma^3-3\gamma^4)h^2
    -4\gamma(19-49\gamma+33\gamma^2-3\gamma^2)h
    -16\gamma^2(1+2\gamma-3\gamma^2)\Big],\nonumber\\
    Q_{yy}^{(3)}(h,\gamma)&=&\frac{3}{4}(1-\gamma)h\Big[ 2h^3
    -2h^2+(1-3\gamma)h+\gamma(1+\gamma)\Big].
\end{eqnarray}
%

\subsubsection{Two-spin expectation value $\langle S_z^2 \rangle$}
\label{app:sub:sub:res_sz2}

%
\begin{eqnarray}
  \label{eq:res_Sz2_sym}
    \frac{4\langle S_z^2\rangle}{N^2}&=&1+\frac{1}{N}\left[
      \frac{P_{zz}^{(1)}(h,\gamma)}{\Xi(h,\gamma)^{1/2}}+2\right]
    +\frac{1}{N^2}\left[
      \frac{P_{zz}^{(2)}(h,\gamma)}{\Xi(h,\gamma)^{2}}
      +\frac{Q_{zz}^{(2)}(h,
        \gamma)}{\Xi(h,\gamma)^{3/2}}\right]\\
    &&+\frac{1}{N^3}(1-\gamma)^2\left[
      \frac{P_{zz}^{(3)}(h,\gamma)}{\Xi(h,\gamma)^{7/2}}
      +\frac{Q_{zz}^{(3)}(h,
        \gamma)}{\Xi(h,\gamma)^{3}}\right]
    +O\left(\frac{1}{N^4}\right),\nonumber
\end{eqnarray}
%
with
%
\begin{eqnarray}
    P_{zz}^{(1)}(h,\gamma)&=&(1+\gamma)-2h=2P_z^{(1)}(h,\gamma),\\
    P_{zz}^{(2)}(h,\gamma)&=&-\frac{1}{4}(1+\gamma)h\Big[16h^2
    +(1+3\gamma)(3+\gamma)\Big]
    +\frac{1}{2}\Big[4h^4+(7+10\gamma+7\gamma^2)h^2+4\gamma^2\Big],\\
    Q_{zz}^{(2)}(h,\gamma)&=&(1+\gamma)(3h^2+\gamma)
    -\frac{1}{2}h\Big[4h^2+3(1+\gamma)^2\Big],\\
    P_{zz}^{(3)}(h,\gamma)&=&\frac{1}{64}(1+\gamma)\Big[ 224h^4
    +(13-202\gamma+13\gamma^2)h^2+16\gamma^2\Big]\\
    &&-\frac{1}{32}h\Big[112h^4+(59-102\gamma+59\gamma^2)h^2
    -2\gamma(19-6\gamma+19\gamma^2)\Big],\nonumber\\
    Q_{zz}^{(3)}(h,\gamma)&=&-\frac{3}{4}(1+\gamma)h(3h^2-\gamma)
    +\frac{3}{4}h^2\Big[4h^2+(1-\gamma)^2\Big].
\end{eqnarray}
%


\subsection{Broken phase}
\label{app:sub:results_broken}

In what follows, we will denote $\Psi(h)=1-h^2$.


\subsubsection{Ground-state energy per spin $e_0(N)$}

%
\begin{equation}
  \label{eq:res_e0_broken}
   e_0(N)=-\frac{1}{2}(1+h^2)
   +\frac{1}{N}\left[(1-\gamma)^{1/2}\Psi(h)^{1/2}
     -\frac{1}{2}(1-\gamma)\right]
   +\frac{1}{N^2}\left[\frac{\gamma h^2-2+\gamma}{2\Psi(h)}
     +\frac{(1-\gamma)^{1/2}}{\Psi(h)^{1/2}}\right]
   +O\left(\frac{1}{N^3}\right).
\end{equation}
%


\subsubsection{Gap $\Delta(N)$}

%
\begin{eqnarray}
  \label{eq:res_Delta_broken}
   \Delta(N)=2(1-\gamma)^{1/2}\Psi(h)^{1/2}
   +\frac{1}{N}\left[-2\frac{(1-2\gamma)h^2+2-\gamma}{\Psi(h)}
     +\frac{2(1-\gamma)^{1/2}}{\Psi(h)^{1/2}}\right]
   +O\left(\frac{1}{N^2}\right).
\end{eqnarray}
%


\subsubsection{One-spin expectation values}

%
\begin{eqnarray}
   \frac{2\langle S_x\rangle}{N}&=&\Psi(h)^{1/2}
   +\frac{1}{N}\left[
     -\frac{(1-2\gamma)h^2+2-\gamma}{2(1-\gamma)^{1/2}\Psi(h)}
   +\frac{h^2}{\Psi(h)^{1/2}}+\Psi(h)^{1/2}\right]
   +O\left(\frac{1}{N^2}\right),\\
   \frac{2\langle S_y\rangle}{N}&=&0
   +O\left(\frac{1}{N^2}\right),\\
   \frac{2\langle S_z\rangle}{N}&=&h
   +\frac{1}{N}\left[\frac{(1-\gamma)^{1/2}h}{\Psi(h)^{1/2}}\right]
   +O\left(\frac{1}{N^2}\right).
\end{eqnarray}
%


\subsubsection{Two-spin expectation values}

%
\begin{eqnarray}
   \frac{4\langle S_x^2\rangle}{N^2}&=&\Psi(h)
   +\frac{1}{N}\left[
     \frac{h^2(h^2-3+2\gamma)}{(1-\gamma)^{1/2}\Psi(h)^{1/2}}
   +2+\frac{(h^2-2+\gamma)\Psi(h)^{1/2}}{(1-\gamma)^{1/2}}\right]
   +O\left(\frac{1}{N^2}\right),\\
   \frac{4\langle S_y^2\rangle}{N^2}&=&\frac{1}{N}
   \frac{\Psi(h)^{1/2}}{(1-\gamma)^{1/2}}
   +O\left(\frac{1}{N^2}\right),\\
   \frac{4\langle S_z^2\rangle}{N^2}&=&h^2
   +\frac{1}{N}\left[\frac{2(1-\gamma)^{1/2}h^2}{\Psi(h)^{1/2}}
     +(1-\gamma)^{1/2}\Psi(h)^{1/2}\right]
   +O\left(\frac{1}{N^2}\right).
\end{eqnarray}
%
\end{widetext}

%
%
%
%


\end{document}